\documentclass[12pt]{article}
\usepackage{eurosym}
\usepackage{amssymb,amsmath,epsfig}
\usepackage{color}
\usepackage{lscape}
\usepackage{amsmath,amsfonts,amssymb}
\usepackage{geometry}
\begin{document}

\title{\bf A New Model of Quintessence Compact Stars in Rastall Theory of Gravity}

\author{G. Abbas$^{1}$ \thanks{abbasg91@yahoo.com} and M. R. Shahzad $^{1}$\thanks{rezwan.chaudhery@gmail.com}\\
$^1$ Department of Mathematics The Islamia University\\ of Bahawalpur,
Bahawalpur, Pakistan.}
\date{}

\maketitle
\begin{abstract}
In the present work, we study a new model of anisotropic compact stars in the regime of Rastall theory. To solve the Rastall field equations we have used the Karori and Barua (KB) ansatz along with the quintessence  dark energy characterized by a parameter $\omega_{q}$ with $-1<\omega_{q}<-\frac{1}{3}$. We present a comparative study to demonstrate the physical acceptance of our proposed model. We compare the numerical values of physical parameters obtained from our model with those of general relativity ($GR$) model given by Bhar \cite{1} and observe that our model is more compatible (for some chosen values of Rastall dimensionless parameter $\gamma=\kappa\lambda$) with observational data than $GR$ model. For this analysis we have consider four different compact stars, $SAX J1808-3658 (SSI)$, $4U 1820-30$, $Vela X-12$ and $PSR J1416-2230$ with radii $7.07km$, $10km$, $9.99km$ and $10.3km$, respectively. In this investigation we also present some physical aspects of the proposed model necessary to check the validity of the model and inferred that our model is acceptable physically and geometrically.
\end{abstract}

\section{Introduction}
%In astrophysics, compact objects such as quark stars and neutrons stars are treated as test beds to analyze certain properties of gravitational fields. In particular, relativistic compact object models have attracted the researchers since last few decades, however Schwarzchild proposed the first exact solution to Einstein's field equations in 1916, for the interior of dense bodies in hydrostatic equilibrium. This provides the direction to the researchers for the learning of an exact solution for astrophysical objects satisfying the variety of criteria which is physically acceptable. In the development of numerous areas of gravitational field like stellar modeling, black hole solution, solar system test, gravitational collapse and so on, the exact solution plays a vital role. Indeed, attaining a regular interior solution for astrophysical compact objects have significant sequels when it comes to solving the field equations. Thus in modern astrophysics, studying the stellar objects from their microscopic composition and properties of dense matter is one of the most fundamental problems. For the construction of theoretical model of stellar objects, innovative work of Volkoff, Tolman, Oppenheimer and Schwarzchild  is of vital importance. Combining the history of relativistic stars, a comprehensive literature can be found by conceiving the certain relativistic effects in the creation of stellar models, where physicists have developed the notion of white dwarf, neutron stars and black holes, representing evolutionary fate of ordinary stars.

To describe the accelerated expansion of our cosmos, a variety of extended theories of gravity have been suggested in last few decades. In this connection an interesting modification of $GR$ has been proposed by Rastall \cite{22} in 1972 , which attracted the attention of many researchers \cite{55}-\cite{46}. In this proposal Rastall confronted the conservation law of energy momentum tensor (i.e., $\nabla_{\nu}T^{\mu\nu}=0$) in a curved space-time, which is the key point of his modification to $GR$. In fact in a curved-space time this conservation law (i.e., $\nabla_{\nu}T^{\mu\nu}=0$) may not hold. Indeed, in Rastall's proposal, covariant divergence of energy momentum tensor is directly proportional to the derivative of scalar curvature, i.e., $\nabla_{\mu}T^{\mu\nu}$$\propto$ $R^{; \nu}$. In this modification Rastall introduced a coupling parameter, whose particular limiting value (i.e., zero coupling) lapse back this modified form to $GR$. Moreover, an other important feature of Rastall theory is that the field equations are quite simpler than the other modified theories and hence easier to investigate.

 Recently this modification to $GR$ has grabbed the renewed attention of researchers because of his interesting behavior in cosmological and astrophysical scenarios. In order to investigate such a new theory (Rastall theory), it is worthwhile to study the construction of astral models also. For example, in this scenario, some rotating and non-rotating black hole solutions have been studied by special arrangement of the parameters in the realm of Rastall theory \cite{23b}-\cite{23e}. Also, the thermodynamics of black hole was discussed in \cite{40}. Fabris et al. \cite{23a} have explored many interesting and new unexpected features of the cosmological models in the realm of Rastall theory. Moradpour and Salako \cite{48} have studied the limitation of Rastall theory due to the application of the Newtonian limit to the theory. They also compare the thermodynamic quantities such as, entropy, energy and work of the system with their counterpart in the context of Einstein theory and observed the better role of Rastall's proposal on the thermodynamics of the system. Properties of the traversable asymptotically flat wormhole solutions were studied in \cite{54} in the Rastall framework. Moradpour et al. \cite{45} observed the remarkable differences between the Rastall theory and $GR$, in cosmological solutions. Moradpour et al. \cite{46} further generalized the Rastall theory, to discuss the cosmic accelerated expansion. They established the non-minimal coupling between the pressureless matter field and geometry, which may be considered as the dark Energy responsible for the present accelerating cosmic phase.

Recently Darabi et al. \cite{47} compare the Rastall theory of gravity with $GR$ and conclude that the claim of Visser \cite{470} that, the equivalence of these two theories, was not correct. They argued that Rastall theory is more general than $GR$. Indeed, Rastall theory is an "open" theory when contrasted with $GR$ and henceforth prepared to confront the challenges of cosmological observations as well as quantum gravity ( for more detailed discussion see, for instance, \cite{47}). Also the influence of the Rastall parameter on static spherically symmetric perfect fluid sphere also been studied by Hansraj et al. in \cite{47a}. They discussed the behavior of well known stellar model proposed by Tolman (R. C. Tolman, Physical Review, \textbf{55}, 364-373 (1939)) in the realm of Rastall theory and compare their findings with $GR$ model and observed that even where $GR$ model shows defective behavior, particular Rastall models fulfil the basic requirements for physically viable model. Hansraj and Banerjee \cite{47b} further studied the influence of linear barotropic equation of state on stellar configuration in Rastall framework. They found a physically acceptable solutions influenced by the Rastall parameter, following the Saslaw's technique to find the unique isothermal fluid solution. In order to become familiar with all the interesting aspects of the Rastall theory, we think that more studies are needed, based on this theory.

Anisotropy refers to the directionally dependent properties of materials. However in the present context anisotropy $\triangle$ may be defined as $\triangle=\frac{2}{r}(p_{t}-p_{r})$, where $p_{t}$, $p_{r}$ and $r$ denoting the transverse pressure, radial pressure and radius of the stellar object, respectively. The effects of anisotropic pressure in the formation of stellar objects structure and their evolution, have been studied by many researchers. Ruderman \cite{2} originally introduced the concept of anisotropy. Later on Bowers and Liang \cite{3} argued that dense matter may be anisotropic because of super-fluxidity and super-conductivity in the presence of complex strong interactions. Herrera and Santos \cite{4} provide an exhaustive discussion to describe the plausible reasons for the occurrence of local anisotropy in self-gravitating systems and describe its consequences. The role of anisotropy on stellar models has been discussed by Maurya et al.\cite{a4} using the functional form of the pressure anisotropy proposed by Lake. Maurya et al.\cite{a5} also suggested that the probability of having anisotropy is considerably higher in compact stars due to the relativistic interaction among the particles and throughout the region, to conserve any uniform motion they become too random. The gravitational collapse and stability of the anisotropic strange stars have also been investigated by Herrera and his collaborators \cite{4a}-\cite{4d}, while in the regime of modified theories of gravity  Sharif and Abbas \cite{5a}-\cite{5c} have studied the related work. Different factors are responsible for the pressure anisotropy , such as several condensate states (like meason condensates, pion condensate etc.) and very high density in the core. Moreover, the basic idea behind the anisotropy and their consequences in different fields can be found in the literature \cite{5}-\cite{17}.

For the present work, motivated from the recent studies in the Rastall framework, we investigate a new anisotropic compact stars model in the presence of quintessence field in the Rastall framework. We study some physical aspects of the presented model in detail to check the physical plausibility and find the realistic behavior of the model. For demonstration we plot the graphs of the physical parameters. Also a comparative study is presented and observed that our model is more compatible with observational data than $GR$ model. The plane of the paper is as follows: in the upcoming section, we formulate the Rastall field equations in the presence of quintessence field and obtain their solutions using Karori and Barua \cite{41} ansatz.  We give the detailed discussion of the physical features of presented model in section \textbf{3}. Finally, we summarize our findings in section \textbf{4}.

\section{Rastall Field Equations in the Presence of Quintessence Field}
In the Rastall's proposal, the energy-momentum conservation law is revised as \cite{22}
\begin{equation}\label{1}
\nabla_{\nu}T^{\nu}_{\mu}=\lambda R_{,\mu},
\end{equation}
where $\lambda$ is a constant called Rastall parameter, which measures the deviation from $GR$ and describes the affinity of the matter field to couple with geometry. According to this modification, the field equations are formulated as
\begin{equation}\label{2}
G_{\mu\nu}+\kappa \lambda g_{\mu\nu} R= \kappa T_{\mu\nu},
\end{equation}
where $\kappa$ represents the gravitational coupling constant of Rastall theory. From (\ref{2}), one can obtain $R(4\kappa\lambda-1)=T$, which shows that $\kappa\lambda=\frac{1}{4}$ is not allowed in this theory, because $T$ is not always zero. Moreover, it is also argued by Rastal that if we use the Newtonian limit and define the Rastall dimensionless parameter $\gamma=\kappa\lambda$, then one can write the $\kappa$ and $\lambda$ in the following form
\begin{equation}\label{3}
\kappa=\frac{4\gamma-1}{6\gamma-1}8\pi,
\end{equation}
\begin{equation}\label{4}
\lambda=\gamma\frac{(6\gamma-1)}{8\pi(4\gamma-1)},
\end{equation}
which show that the Einstein result ($\kappa=8\pi$) can be obtained using an appropriate limit $\lambda=0$ which is equivalent to $\gamma=0$ limit. Note that from equation (\ref{3}), for $\gamma=\frac{1}{6}$, the Rastall coupling constant diverges, thus $\gamma=\frac{1}{6}$ is not allowed in this theory. Similarly it is evident from equation (\ref{4}), $\gamma=\frac{1}{4}$ is also not allowed \cite{48}. Thus, finally the Rastall field equations can be formulated as
\begin{equation}\label{5}
G_{\mu\nu}+\gamma g_{\mu\nu} R=\frac{4\gamma-1}{6\gamma-1} 8\pi T_{\mu\nu}.
\end{equation}
Thus Newtonian limit represents that, indeed, both the cases $\gamma=\frac{1}{6}$ and $\gamma=\frac{1}{4}$ are not allowed in this theory.
Now in the presence of quintessence dark energy, the Rastall field equations formulated in equation (\ref{5}) can be written as
\begin{equation}\label{6}
G_{\mu\nu}+\gamma g_{\mu\nu} R=\frac{4\gamma-1}{6\gamma-1} 8\pi (T_{\mu\nu}+\tau_{\mu\nu}),
\end{equation}
where $\tau_{\mu\nu}$ stands for the energy momentum tensor of the quintessence like field with characteristic parameter $\omega_{q}$ satisfying the constraints $-1<\omega_{q}<-\frac{1}{3}$. The components of $\tau_{\mu\nu}$ are $\tau_{t}^{t}=\tau_{r}^{r}$=-$\rho_{q}$ and $\tau_{\theta}^{\theta}=\tau_{\phi}^{\phi}=\frac{\rho_{q}}{2}(3\omega_{q}+1)$ as defined by Bhar \cite{1}.
For static spherically symmetric spacetime, the line element is given by
\begin{equation}\label{r1}
ds^2=-e^{\nu(r)}dt^{2}+e^{\mu(r)}dr^{2}+r^2(d\theta^2+\sin^2\theta{d\phi^2}),
\end{equation}
 here we chose $\mu(r)=Ar^{2}$ and $\nu(r)=Br^{2}+C$ as given in \cite{41}. The anisotropic and static matter configuration is obtained by
\begin{equation}\nonumber
T_{\sigma\delta}=(\rho+p_{t})\xi_{\sigma}\xi_{\delta}+p_{t}g_{\sigma\delta}+(p_{r}-p_{t})\eta_{\sigma}\eta_{\delta},
\end{equation}
where $\rho$, $p_{r}$ and $p_{t}$ representing respectively energy density, radial and transverse pressures observed by a comoving observer. The $\xi^{\sigma}$ representing four-velocity and $\eta^{\sigma}$ is its unit normal vector, such that the constraints $\xi^{\sigma}=e^{-\frac{\nu}{2}}\delta^{\sigma}_{0}$, $\xi^{\sigma}\xi_{\sigma}=-1$ and $\eta^{\sigma}=e^{-\frac{\mu}{2}}\delta^{\sigma}_{1}$, $\eta^{\sigma}\eta_{\sigma}=1$ are satisfied. For metric (\ref{r1}), the Rastall field equations (\ref{6}) are formulated as:

\begin{equation}\label{r2}
8\pi(\rho+\rho_{q})\bigg(\frac{4\gamma-1}{6\gamma-1}\bigg)=e^{-\mu}\bigg[\frac{\mu^{'}r+e^{\mu}-1}{r^2}+\gamma \bigg\{\nu''+\nu'^2-\mu'\nu'-\frac{2}{r}(\mu'-\nu')-\frac{2}{r^2}(e^\mu-1)\bigg\}\bigg],
\end{equation}

\begin{equation}\label{r3}
8\pi (p_{r}-\rho_{q})\bigg(\frac{4\gamma-1}{6\gamma-1}\bigg)=e^{-\mu}\bigg[\frac{\nu^{'}r-e^{\mu}+1}{r^2}-\gamma \bigg\{\nu''+\nu'^2-\mu'\nu'-\frac{2}{r}(\mu'-\nu')-\frac{2}{r^2}(e^\mu-1)\bigg\}\bigg],
\end{equation}

\begin{eqnarray}\nonumber
8\pi (p_{t}+\frac{(3\omega_{q}+1)\rho_{q}}{2})\bigg(\frac{4\gamma-1}{6\gamma-1}\bigg)=&&e^{-\mu}\bigg[\frac{-\mu'\nu'r+\nu'^2 r+2 \nu'' r-2\mu'+2\nu'}{4 r}-\gamma\{\nu''
\\&&+\nu'^2-\mu'\nu' -\frac{2}{r}(\mu'-\nu')-\frac{2}{r^2}(e^\mu-1)\}\bigg]\label{r4}.
\end{eqnarray}
In the above system of equations, there are three equations (\ref{r2})-(\ref{r4}) with four unknowns viz. $\rho$, $p_{r}$, $p_{t}$ and $\rho_{q}$. Thus, in order to solve the above system we consider that the radial pressure $p_{r}$ is proportional to the matter density $\rho$, i.e.,
\begin{equation}\label{7}
p_{r}=\beta\rho,
\end{equation}
where $\beta$ involved in the above equation represents the equation of state parameter satisfying $0<\beta<1$.
Now, using eqs. (\ref{r2})-(\ref{r4}) and (\ref{7}) along with metric (\ref{r1}), one can easily obtain
\begin{equation}\label{r5}
\rho=\bigg(\frac{A+B}{4\pi (1+\beta)}\bigg)\bigg(\frac{6\gamma-1}{4\gamma-1}\bigg)e^{-Ar^{2}},
\end{equation}

\begin{equation}\label{r6}
p_{r}=\bigg(\frac{A+B}{4\pi (1+\beta)}\bigg)\bigg(\frac{6\gamma-1}{4\gamma-1}\bigg)\beta e^{-Ar^{2}},
\end{equation}

\begin{eqnarray}\nonumber
\rho_{q}&&=\bigg(\frac{e^{-Ar^{2}}}{8\pi}\bigg)\bigg(\frac{6\gamma-1}{4\gamma-1}\bigg)\bigg[2A-\frac{1}{r^{2}}+\frac{e^{Ar^{2}}}{r^{2}}-\frac{2(A+B)}{1+\beta}
\\&&-\gamma\bigg\{4 (A-B)(1+Br^{2})-2(B-\frac{e^{Ar^{2}}-1} {r^2})\bigg\}\bigg]\label{r7},
\end{eqnarray}

\begin{eqnarray}\nonumber
p_{t}&&=\bigg(\frac{e^{-Ar^{2}}}{8\pi}\bigg)\bigg(\frac{6\gamma-1}{4\gamma-1}\bigg)\bigg[B-(A-B)(1+Br^{2})+\gamma\bigg(4(A-B)
\\\nonumber&&\times(1+Br^{2})-2(B-\frac{e^{Ar^{2}}-1}{r^2})\bigg)-\bigg(\frac{3\omega+1}{2}\bigg)\bigg\{2A-\frac{1}{r^{2}}+\frac{e^{Ar^{2}}}{r^{2}}
\\&&-\frac{2(A+B)}{1+\beta}-\gamma\bigg(4 (A-B)(1+Br^{2})-2(B-\frac{e^{Ar^{2}}-1} {r^2})\bigg)\bigg\}\bigg]\label{r07}.
\end{eqnarray}

Now, to determine the values of unknown constants involved in metric (\ref{r1}), we match our interior metric (\ref{r1}) with the appropriate exterior metric, i.e., Schwarzschild metric
\begin{equation}\label{a1}
ds^2=-\bigg(1-\frac{2M}{r}\bigg)dt^{2}+\frac{dr^{2}}{1-\frac{2M}{r}}+r^2(d\theta^2+\sin^2\theta{d\phi^2}),
\end{equation}
 where $M$ represents the mass of the compact object (i.e., mass of black hole). The continuity of the metric coefficients i.e., $g_{ii}$, $i=r,t$ and $\frac{\partial g_{tt} }{\partial r}$ at the boundary $r=R$ produces the following equations
 \begin{equation}\label{a2}
1-\frac{2M}{R}=e^{BR^{2}+C},
\end{equation}

 \begin{equation}\label{a3}
\frac{1}{1-\frac{2M}{R}}=e^{AR^{2}},
\end{equation}

 \begin{equation}\label{a4}
\frac{M}{R^{}2}=BRe^{BR^{2}+C}.
\end{equation}
Using the above Eqs. (\ref{a2})-(\ref{a4}), one can easily obtain,

 \begin{equation}\label{a5}
A=-\frac{1}{R^{2}}ln\bigg(1-\frac{2M}{R}\bigg),
\end{equation}

 \begin{equation}\label{a6}
B=\frac{M}{R^{3}\bigg(1-\frac{2M}{R}\bigg)},
\end{equation}

 \begin{equation}\label{a7}
C=ln\bigg(1-\frac{2M}{R}\bigg)-\frac{M}{R\bigg(1-\frac{2M}{R}\bigg)}.
\end{equation}
Thus using the above expressions, the values of $A$ and $B$ are given in \textbf{Table-1}, for different four compact stars.

\begin{table}[ht]
\caption{Calculated values of $A$ and $B$}
\begin{center}
\begin{tabular}{|c|c|c|c|c|}
\hline {Compact Star}&  \textbf{ $M(M_\odot)$} & \textbf{$R(km)$}  &\textbf{ $A(km ^{-2})$}& \textbf{$B(km ^{-2})$}
\\\hline $SAX J 1808.4-3658 (SSI)$  & 1.435  & 7.07     &0.01823156974    & $0.014880115690$
\\\hline $4U 1820-30$               & 2.25   & 10.0     &0.01090644119    & $0.009880952381$
\\\hline $Vela X-12$                & 1.77   & 9.99     &0.00741034129    & $0.005485958565$
\\\hline $PSR J1614-2230$           & 1.97   & 10.3     &0.00782944033    & $0.006102145623$
\\\hline
\end{tabular}
\end{center}
\end{table}

\section{Physical aspects and comparative study of the physical parameters }
In this section, we inspect more details about the configuration of stellar object by propounding some analytical calculations and concentrated on physical properties of stellar interior. In this connection physical features of the presented model are also observed by graphical representation of obtained solutions and finally, we compare our results with $GR$ model and the observational constraints as well.
\subsection{Evolution of  pressure, energy density and quintessence field}
For a realistic model, density and pressure (in the present case $p_{t}$ and $p_{r}$) should be positive inside the stellar body, i.e., $\rho>0$,  $p_{t}>0$ and $p_{r}>0$. Moreover these parameters should be (i) regular throughout the configuration of the stellar object (ii) acquired their maximum values at the center and (iii) decreasing functions of $r$ from center to surface. In our present model to fulfill these basic conditions, we propound the constraints on the Rastall dimensionless parameter $\gamma$, as $\gamma>\frac{1}{4}$ and $\gamma<\frac{1}{6}$.

Now, using Eqs. (\ref{r5}) and (\ref{r6}), one can easily obtain the following equations

\begin{eqnarray}\label{r10}
\frac{d\rho}{dr}=-\bigg(\frac{Ar(A+B)}{2\pi (1+\beta)}\bigg)\bigg(\frac{6\gamma-1}{4\gamma-1}\bigg)e^{-Ar^{2}},
\end{eqnarray}

\begin{eqnarray}\label{r11}
\frac{dp_{r}}{dr}=-\bigg(\frac{Ar\beta(A+B)}{2\pi (1+\beta)}\bigg)\bigg(\frac{6\gamma-1}{4\gamma-1}\bigg)e^{-Ar^{2}},
\end{eqnarray}
\begin{eqnarray}\label{r12}
\frac{d^{2}\rho}{dr^{2}}=\bigg(\frac{A(A+B)}{\pi (1+\beta)}\bigg)\bigg(\frac{6\gamma-1}{4\gamma-1}\bigg)e^{-Ar^{2}}(Ar^{2}-\frac{1}{2}),
\end{eqnarray}
\begin{eqnarray}\label{r13}
\frac{d^{2}p_{r}}{dr^{2}}=\bigg(\frac{A\beta(A+B)}{\pi (1+\beta)}\bigg)\bigg(\frac{6\gamma-1}{4\gamma-1}\bigg)e^{-Ar^{2}}(Ar^{2}-\frac{1}{2}).
\end{eqnarray}
It is evident from eqs. (\ref{r10})-(\ref{r13}), at $r=0$

\begin{eqnarray}\nonumber
\frac{d\rho}{dr}=0,~~~~~    \frac{dp_{r}}{dr}=0,~~~~~\frac{d^{2}\rho}{dr^{2}}<0,~~~   and          ~~~~~\frac{d^{2}p_{r}}{dr^{2}}<0.
\end{eqnarray}

which ensure that central density as well as central pressure are maximum, as expected due to the decreasing nature of these functions from center to surface. For instance, using the equations (\ref{r5}) and (\ref{r6}), one can obtain the expressions for density at the center and surface and for central pressure respectively as:
 \begin{equation}\label{s1}
(\rho)_{r=0}=\bigg(\frac{A+B}{4\pi (1+\beta)}\bigg)\bigg(\frac{6\gamma-1}{4\gamma-1}\bigg),
\end{equation}

\begin{equation}\label{s2}
(\rho)_{r=R}=\bigg(\frac{A+B}{4\pi (1+\beta)}\bigg)\bigg(\frac{6\gamma-1}{4\gamma-1}\bigg)e^{-AR^{2}},
\end{equation}
\begin{equation}\label{s3}
(p_{r})_{r=0}=\bigg(\frac{A+B}{4\pi (1+\beta)}\bigg)\bigg(\frac{6\gamma-1}{4\gamma-1}\bigg)\beta .
\end{equation}
Using the above expressions, the numerical values of matter density at the center as well as at the surface and central pressure has been calculated and exhibited in \textbf{Tables 2-5}. The same nature as we have investigated analytically of the matter density and pressure (radial and tangential) can eminently be observed from \textbf{Figs. 1 }(upper left panel) and \textbf{3} (first row). The evolution of the quintessence field ($\rho_{q}$) is shown in \textbf{Fig.} (lower right panel). We would like to mention that through out the study, the parameters $\beta$ and ${w}_q$ have been taken as 0.11 and -0.7, respectively.

\subsection{Energy conditions}
 For any physically adequate anisotropic fluid configuration the energy conditions such as null energy condition $(NEC)$, weak energy condition $(WEC$), strong energy condition $(SEC)$ and dominant energy condition (DEC) must be satisfied by the matter inside it. Basically all these energy conditions demand that the energy can never be negative because the negative energy would never make the stable configuration. These energy conditions are fulfilled if the following inequalities are satisfied simultaneously:
\begin{equation}\label{r13a}
NEC: \rho\geq0,
\end{equation}
\begin{equation}\label{r13b}
WEC: \rho +p_{r}\geq0, \rho +p_{t}\geq0,
\end{equation}
\begin{equation}\label{r31c}
SEC:\rho +p_{r}\geq0, \rho +p_{t}\geq0, \rho +p_{r}+2p_{t}\geq0,
\end{equation}
\begin{equation}\label{r13d}
DEC: \rho \geq |p_{r}|, \rho \geq |p_{t}|.
\end{equation}
 We interpret all these energy conditions by graphical analysis. It is clear eminently from \textbf{Figs. 1} that all the energy conditions mentioned above are satisfied for our system.

\subsection{Stability}
Using the radial and transverse sound speeds, formulated below, we examine the stability of the present model.
\begin{equation}\label{r21}
\nu^{2}_{sr}=\beta,
\end{equation}
and
\begin{eqnarray}\nonumber
\nu^{2}_{st}&&= \frac{1}{4Ar^{4}(A+B)}\bigg(e^{Ar^{2}}(1+\beta)(-1-3\omega+6\gamma(1+\omega))+(1+\beta)
\\\nonumber&&\times(1+3\omega-6\gamma(1+\omega)+2B^{2}r^{4}(6\gamma(1+\omega_{q})-1))+2A^{2}r^{4}(-1-2\beta
\\\nonumber&&-3\omega_{q}\beta+6\gamma(1+\omega_{q})(1+\beta)+Br^{2}(1+\beta)(6\gamma(1+\omega_{q})-1))+Ar^{2}
\\\nonumber&&\times\{-2B^{2}r^{4}(1+\beta)(6\gamma(1+\omega_{q})-1)-(1+\beta)(6\gamma(1+\omega_{q})-1-3\omega_{q})
\\&&+2Br^{2}(4+3\omega_{q}+3\beta-15\gamma(1+\omega_{q})(1+\beta))\}\bigg)\label{r22},
\end{eqnarray}
\begin{equation}\label{r23}
\nu^{2}_{st}-\nu^{2}_{sr}=\nu^{2}_{st}-\beta.
\end{equation}
For a physically suitable fluid distribution, radial and transverse sound speeds must be less than the speed of light ( in the present case we adopt the suitable units such that $c=1$, where c is the speed of light) i.e., $0\leq \nu^{2}_{sr} \leq1$, $0\leq \nu^{2}_{sr} \leq1$. This condition for sound speed is called causality condition propounded by Herrera \cite{42}. Equation (\ref{r21}) and the profile of $\nu^{2}_{st}$ presented in \textbf{Figs. 2}(left panel) eminently show that our model meets the causality conditions mentioned above and hence represents the stable configuration. Moreover, according to Herrera and Andreasson \cite{42,43} a region is potentially stable if  $|\nu^{2}_{st}-\nu^{2}_{sr}|\leq1$ is satisfied. Observing from \textbf{Fig. 2} (right panel) one can infer that the above condition for potentially stable region is satisfied for our presented model and hence the system is potentially stable within the entire configuration of the compact star considered here.
\subsection{Measure of Anisotropy}
To measure the anisotropy for the present model, using the relation $\Delta=\frac{2}{r}(p_{t}-p_{r})$ we acquire the expression as follows:
\begin{equation}\label{13}
\Delta=\frac{e^{-Ar^{2}}(6\gamma-1)g_{2}(r)}{4\pi r(4\gamma-1)},
\end{equation}
where
\begin{eqnarray}\nonumber
g_{2}(r)&&=Br^{2}(A-B)(6\gamma(1+\omega_{q})-1)+\frac{(6\gamma(1+\omega_{q})-1-3\omega_{q})(e^{Ar^{2}}-1)}{2r^{2}}
\\\nonumber&&+\frac{3B(1+\omega_{q})-A(1+\beta(4+3\omega_{q}))+3\gamma(1+\beta)(1+\omega_{q})(2A-3B)}{1+\beta}.
\end{eqnarray}
For $\Delta>0$, i.e., $p_{t}>p_{r}$, the anisotropic pressure is outward directed, and for $\Delta<0$, i.e., $p_{t}<p_{r}$, it is directed inward. For better understanding about the anisotropy for the present model we examine its nature graphically. \textbf{Fig. 3} (lower left panel) shows that $\Delta>0$ i.e., $p_{t}>p_{r}$, depicting the repulsive nature of the anisotropic pressure and hence allow the formation of more compact configuration.

\subsection{Equation of state ($EoS$)}
For our system we formulate  the equation of state parameters $w_{r}$ and $w_{t}$, corresponding to normal and transverse directions as follows:
\begin{equation}\label{r8}
w_{r}=\frac{p_{r}}{\rho}=\beta,
\end{equation}
\begin{eqnarray}\label{r9}
w_{t}=\frac{p_{t}}{\rho}&&=\frac{(1+\beta)g_{1}(r)}{2(A+B)},
\end{eqnarray}
where
\begin{eqnarray}\nonumber
g_{1}(r)&&=-2A+2B-3A\omega_{q}+\frac{(A+B)(1+3\omega_{q})}{1+\beta}+3\gamma(2A-3B)(1+\omega_{q})
\\\nonumber&&+Br^{2}(A-B)(6\gamma(1+\omega_{q})-1)+\frac{(6\gamma(1+\omega_{q})-1-3\omega_{q})(e^{Ar^{2}}-1)}{2r^{2}}.
\end{eqnarray}
The evolution of these parameters can be perceived from \textbf{Fig. 4} (see equation (\ref{r8}) for $w_{r}$). It is evident that the values of equation of state parameters $w_{r}$ and $w_{t}$ remains positive throughout the entire region of the fluid sphere and lie in the interval $(0,1)$ which ensure the non-exotic configuration of the stellar object.
\begin{figure}

\end{figure}
\subsection{ Mass function}
The mass function of the compact object with radius $r$ is defined by
\begin{equation}\nonumber
M=4\pi\int_{0}^{r}\rho r^{2}dr=\frac{(A+B)(6\gamma-1)}{2A(1+\beta)(4\gamma-1)}\bigg[\frac{\sqrt{\pi}}{2\sqrt{A}}erf(\sqrt{A}R)-Re^{-Ar^{2}}\bigg].
\end{equation}
The profile of the mass function is exhibited in \textbf{fig. 4 } (first row right panel). It is evident from the plot of $M(r)$ that $M(r)>0$ throughout the stellar interior. Moreover, as $r\rightarrow0$, $M(r)\rightarrow0$ depicting the regularity of the mass function at the center. The numerical values of the mass for different stars considered in this study have been calculated using our proposed model and presented in \textbf{Tables 2-5}. We have also compared these values of mass with that of obtained from $GR$ model and with observational data. It can be seen from \textbf{Tables 2-5} that the numerical values of the masses calculated from the present model are more closer to the observed masses than that of calculated from $GR$ model \cite{1}.

\subsection{ Mass-radius ratio}
This section is concerned about the discussion of mass to radius ratio. The maximum allowable limit for mass to radius ratio is discussed by Bhuchdahl \cite{55a}. He suggested that for a static spherical symmetric isotropic configuration of stellar objects, mass to radius ratio should satisfy the constraint $\frac{2M}{R}<\frac{8}{9}$. Mak and Harko \cite{55b} further generalized the result of Bhuchdahl. We have calculated mass to radius ratio from our model for different strange stars (given in \textbf{Table-1} ) and observed that the Bhuchdahl condition is satisfied (see \textbf{Tables 6-9}) for all the compact stars considered in this study.

Now we define the compactness of the stellar model for our system as the ratio of mass ($M$) and radius ($R$)
\begin{eqnarray}\nonumber
u=\frac{M}{R}=\frac{(A+B)(6\gamma-1)}{2A(1+\beta)(4\gamma-1)}\bigg[\frac{\sqrt{\pi}}{2R\sqrt{A}}erf(\sqrt{A}R)-e^{-AR^{2}}\bigg].
\end{eqnarray}
The profile of the compactification factor $u(R)$ is displayed in \textbf{fig. 12}. The numerical values of the compactification factor$u(R)$ calculated from our model are given in \textbf{Tables 6-9}, for differen compact stars.
\subsection{Surface Redshift}
Using the above compactification factor $u(R)$, the surface redshift $Z_{s}(r)$ is given by
\begin{equation}\label{r251}
Z_{s}=[1-2u]^{-\frac{1}{2}}-1=\bigg[1-\frac{(A+B)(6\gamma-1)(\frac{\sqrt{\pi}erf(\sqrt{A}R)}{2R\sqrt{A}}-e^{-AR^{2}})}{A(1+\beta)(4\gamma-1)}\bigg]^{-\frac{1}{2}}-1.
\end{equation}
The behavior of the surface redshift $Z_{s}$ against the radial coordinate $r$ is shown in \textbf{Fig. 4} (second row right panel). It is evident from the figure that $Z_{s}$ vanishes at the center and gradually increases with the increase in $r$. We have calculated the numerical values of the maximum surface redshift for different astral objects, shown in \textbf{Tables 6-9}. In this connection it is reasonable to discuss here that the upper limit for the surface redshift is suggested as $Z_{s}\leq2$ in \cite{55a,55d,55e} for the isotropic configuration without cosmological constant. However for an anisotropic configuration with cosmological constant, it is proposed that $Z_{s}\leq5$ by Bohmer and Harko \cite{55f}, whereas Ivanov \cite{55g} suggested that $Z_{s}\leq5.211$. In the light of above discussion, for an anisotropic compact star in the absence of cosmological constant the maximum value for the surface redshift calculated using the present model $Z_{s}=0.733$ is in good concurrence.

\begin{figure}
\includegraphics[width=.49\linewidth, height=1.9in]{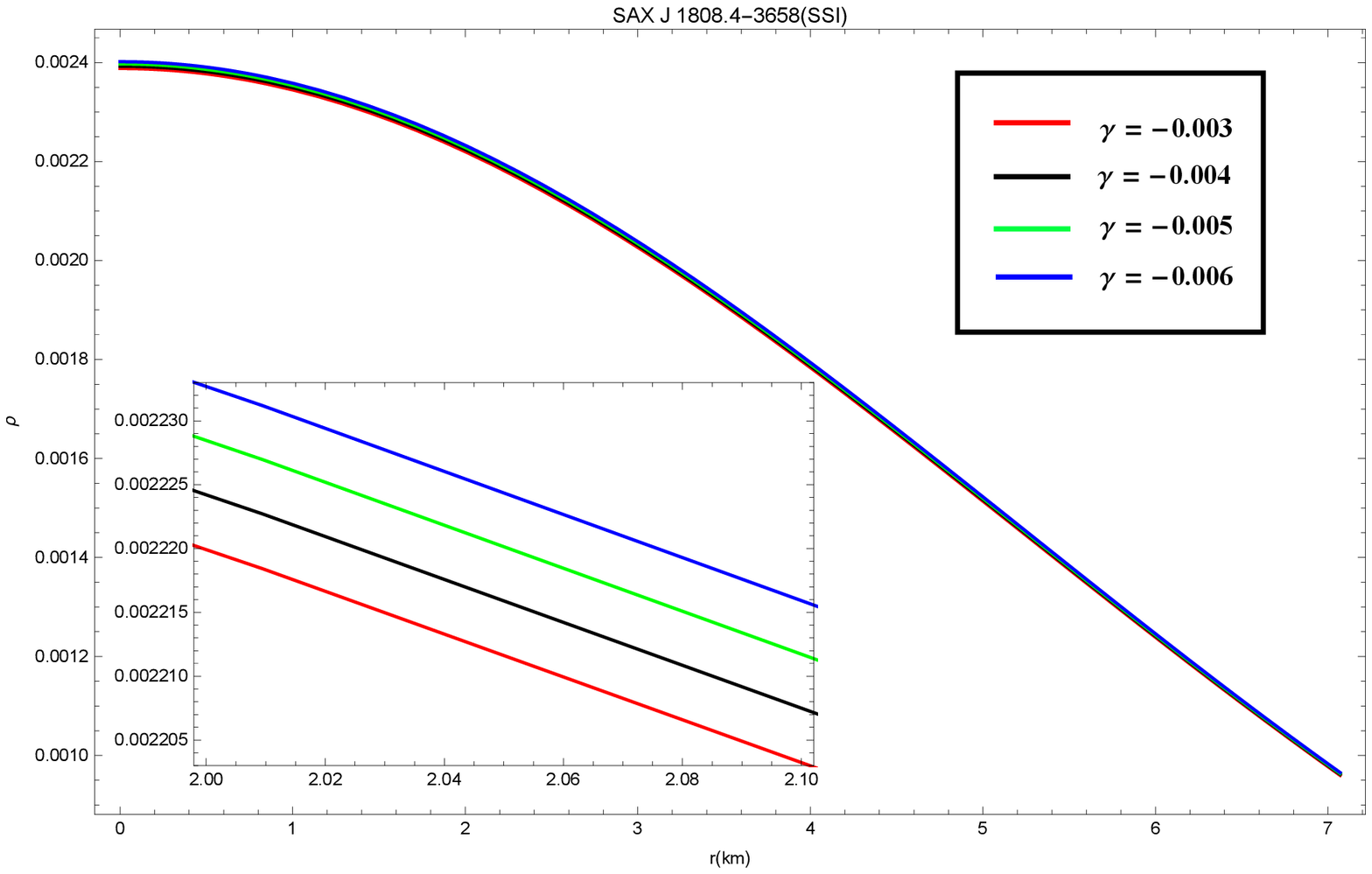}
\includegraphics[width=.49\linewidth, height=1.9in]{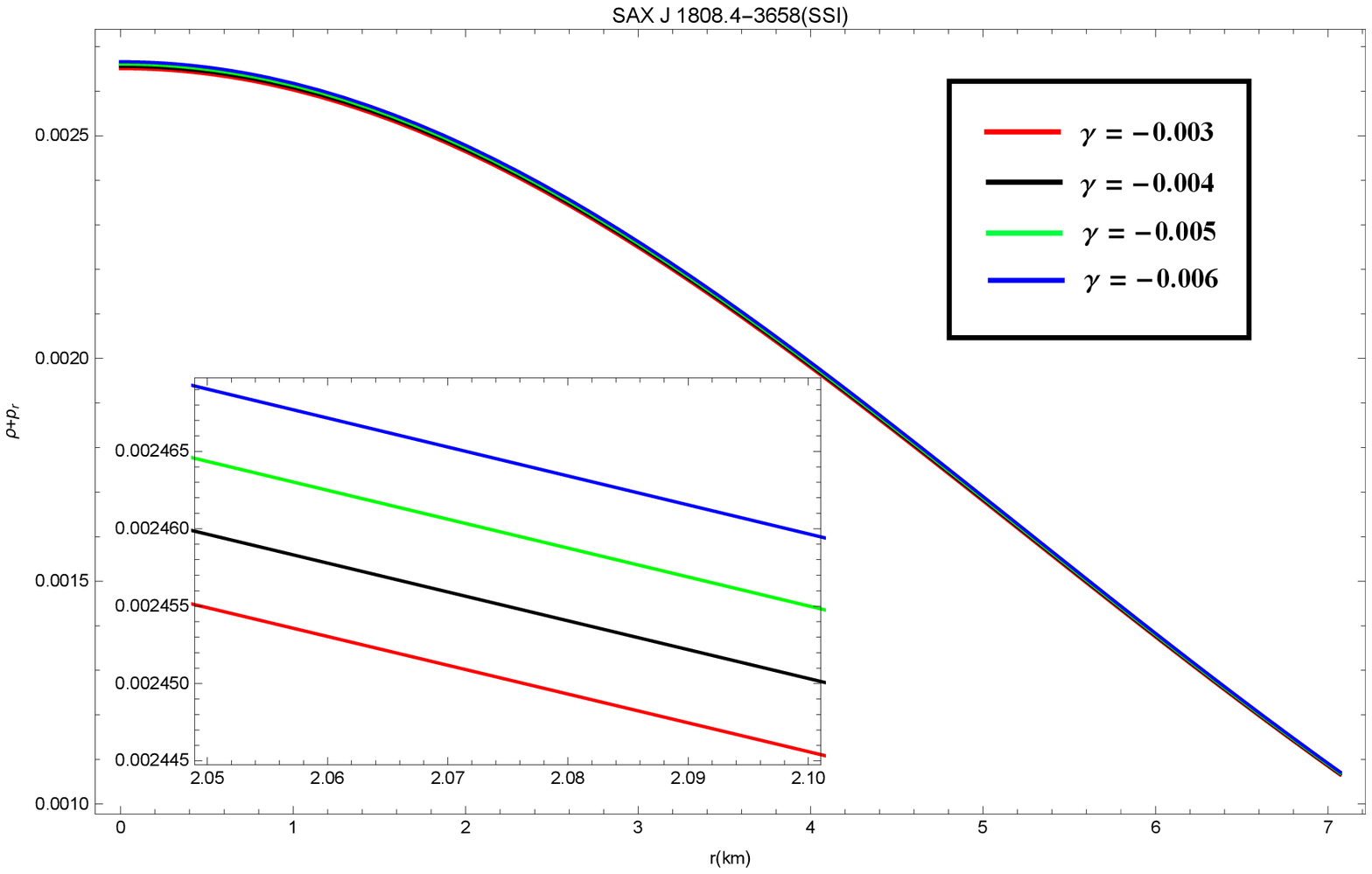}
\includegraphics[width=.49\linewidth, height=1.9in]{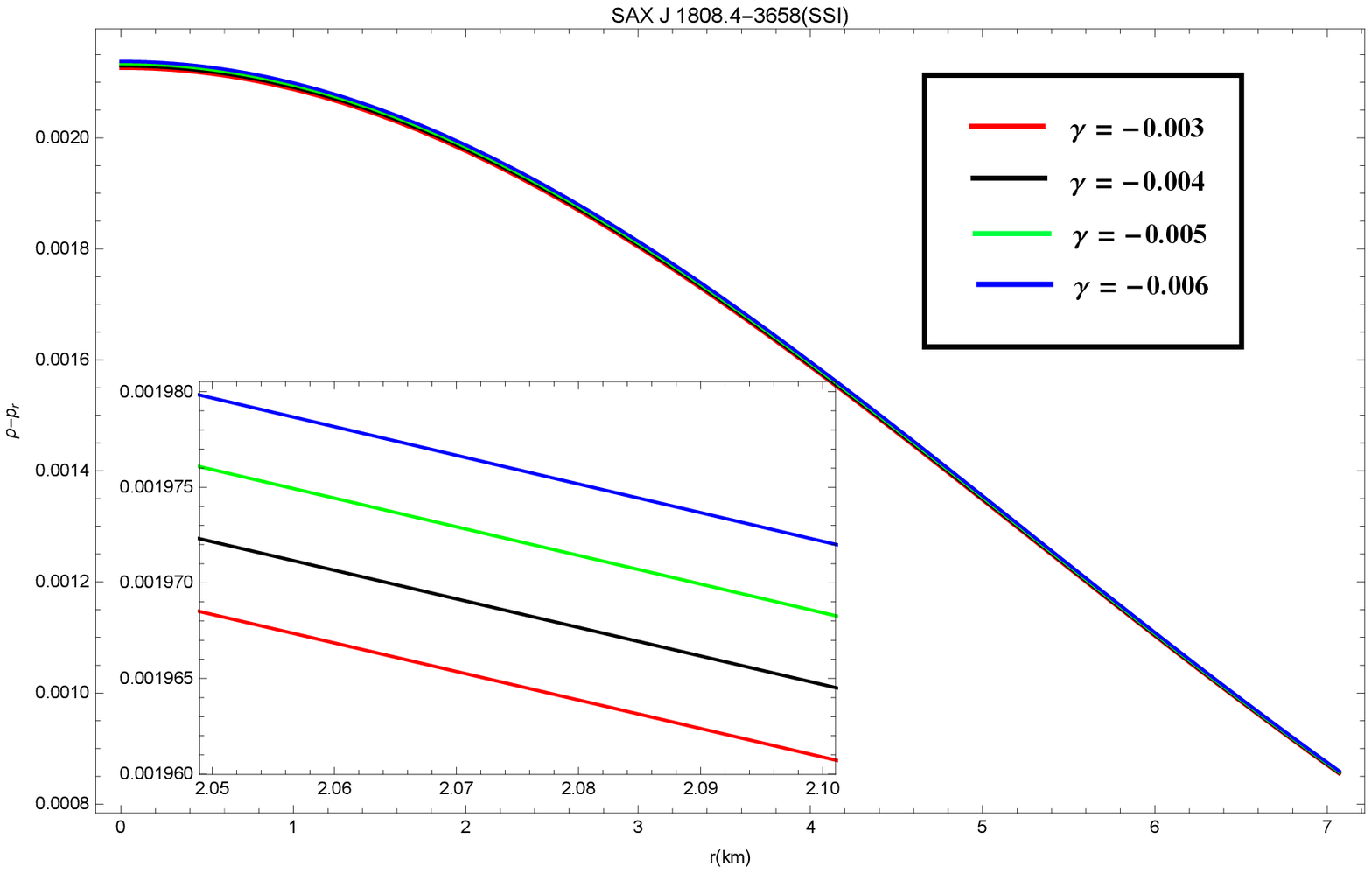}
\includegraphics[width=.49\linewidth, height=1.9in]{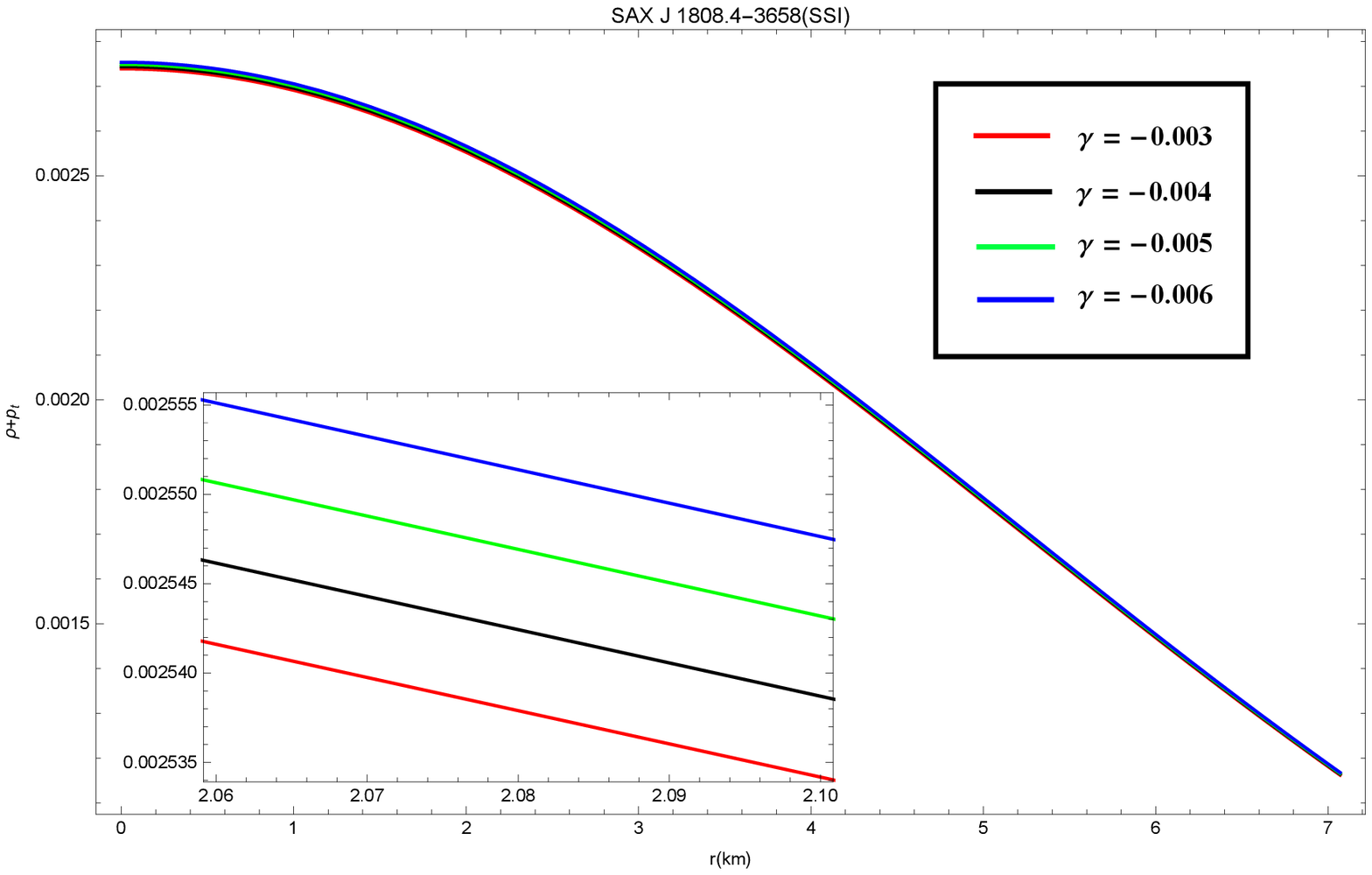}
\includegraphics[width=.49\linewidth, height=1.9in]{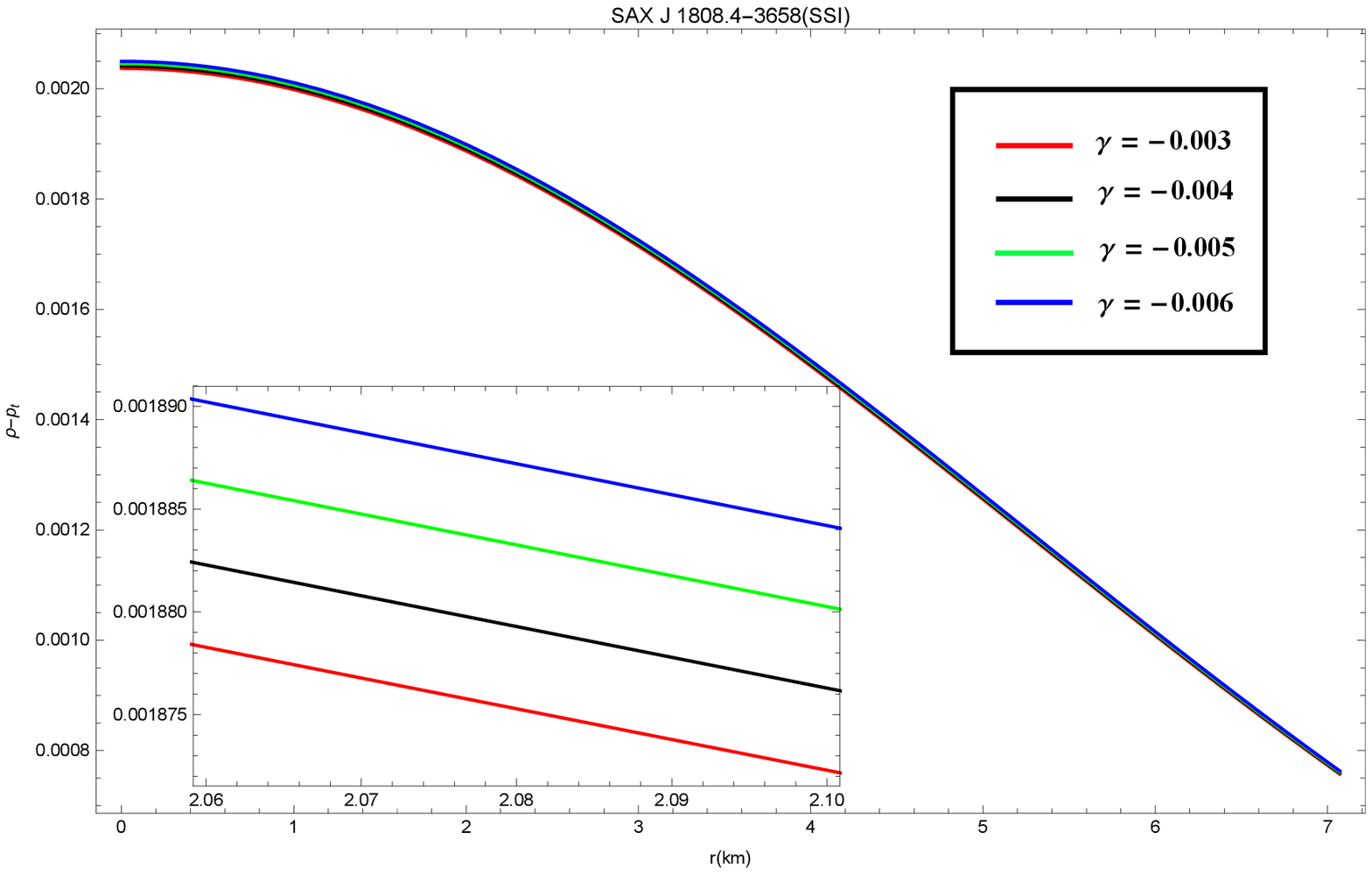}
\includegraphics[width=.49\linewidth, height=1.9in]{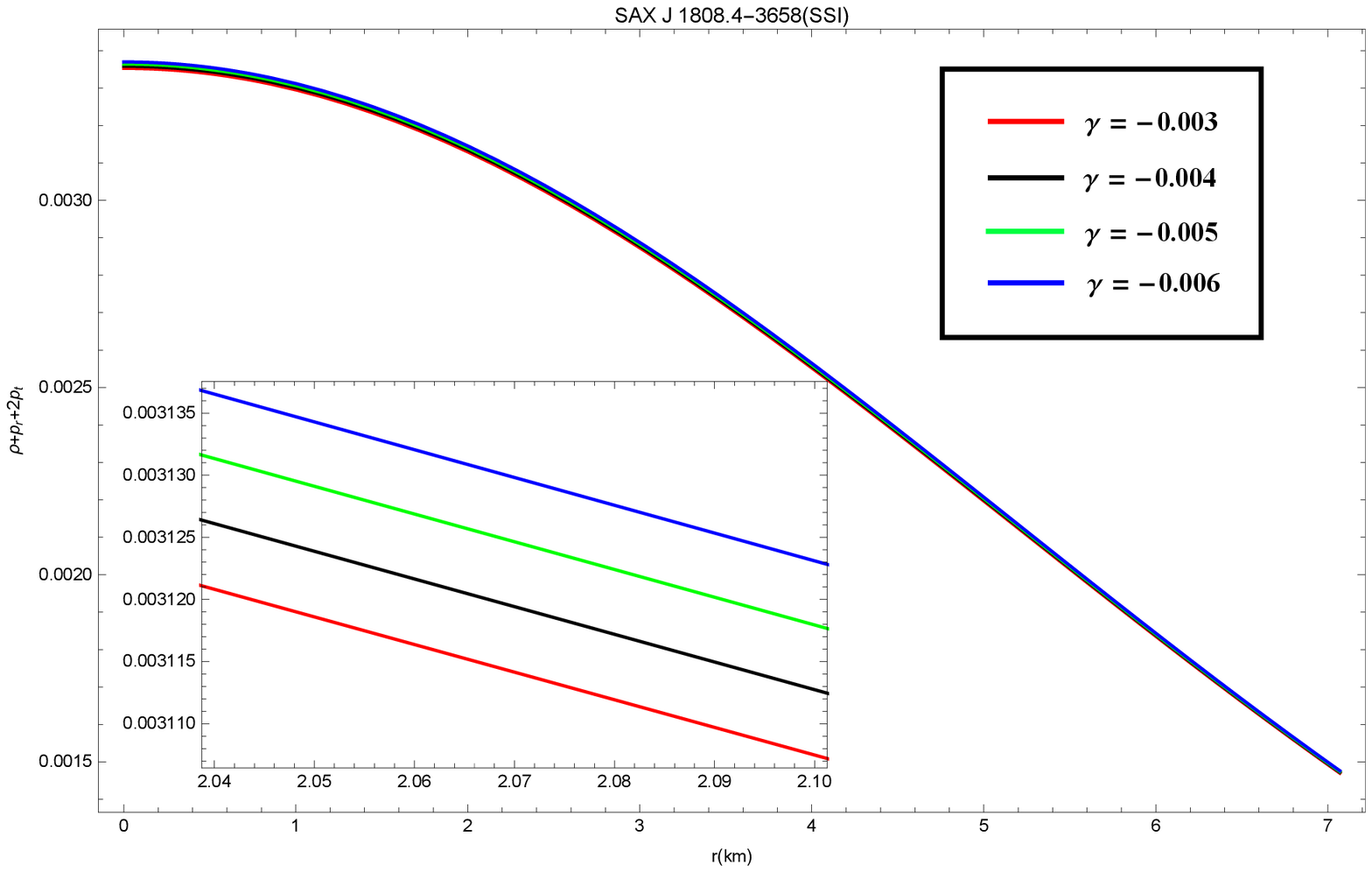}
\caption{Evolution of energy conditions. }
\end{figure}

 \begin{figure}
\includegraphics[width=.49\linewidth, height=1.9in]{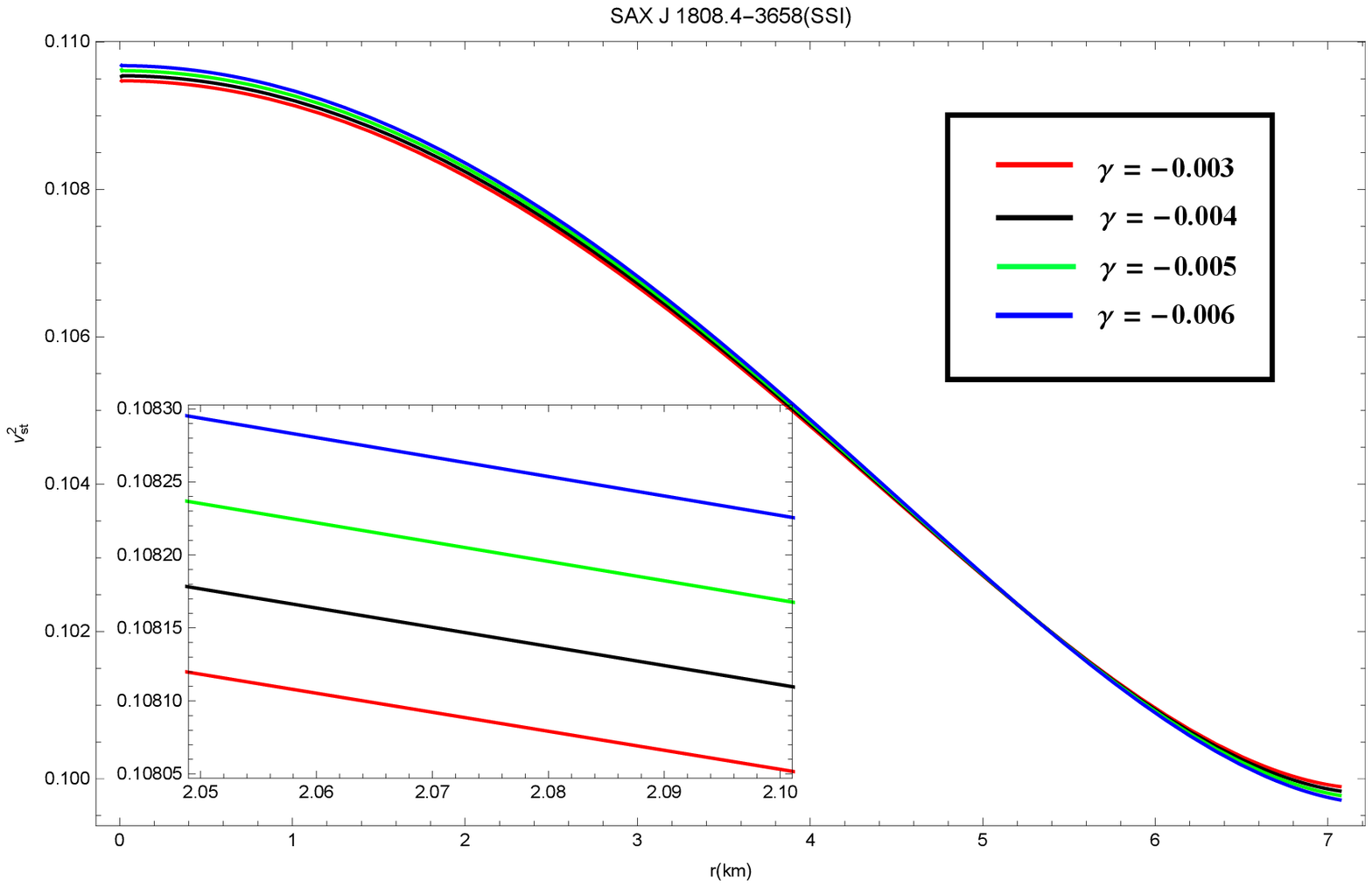}
\includegraphics[width=.49\linewidth, height=1.9in]{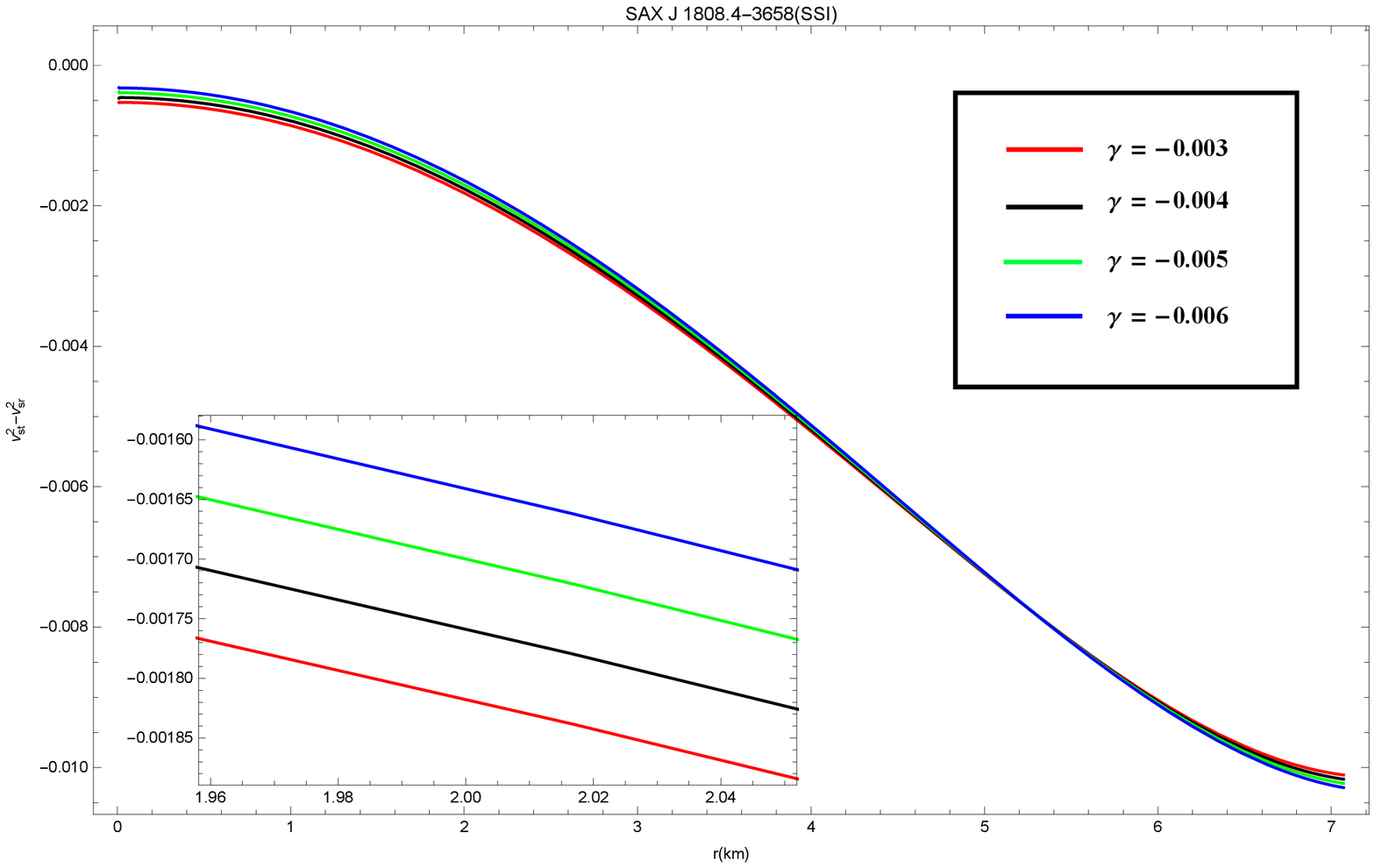}
\caption{Profile of $\nu_{st}^{2}$ and $\nu_{st}^{2}-\nu_{sr}^{2}$. }
\end{figure}

\begin{figure}
\includegraphics[width=.50\linewidth, height=1.8in]{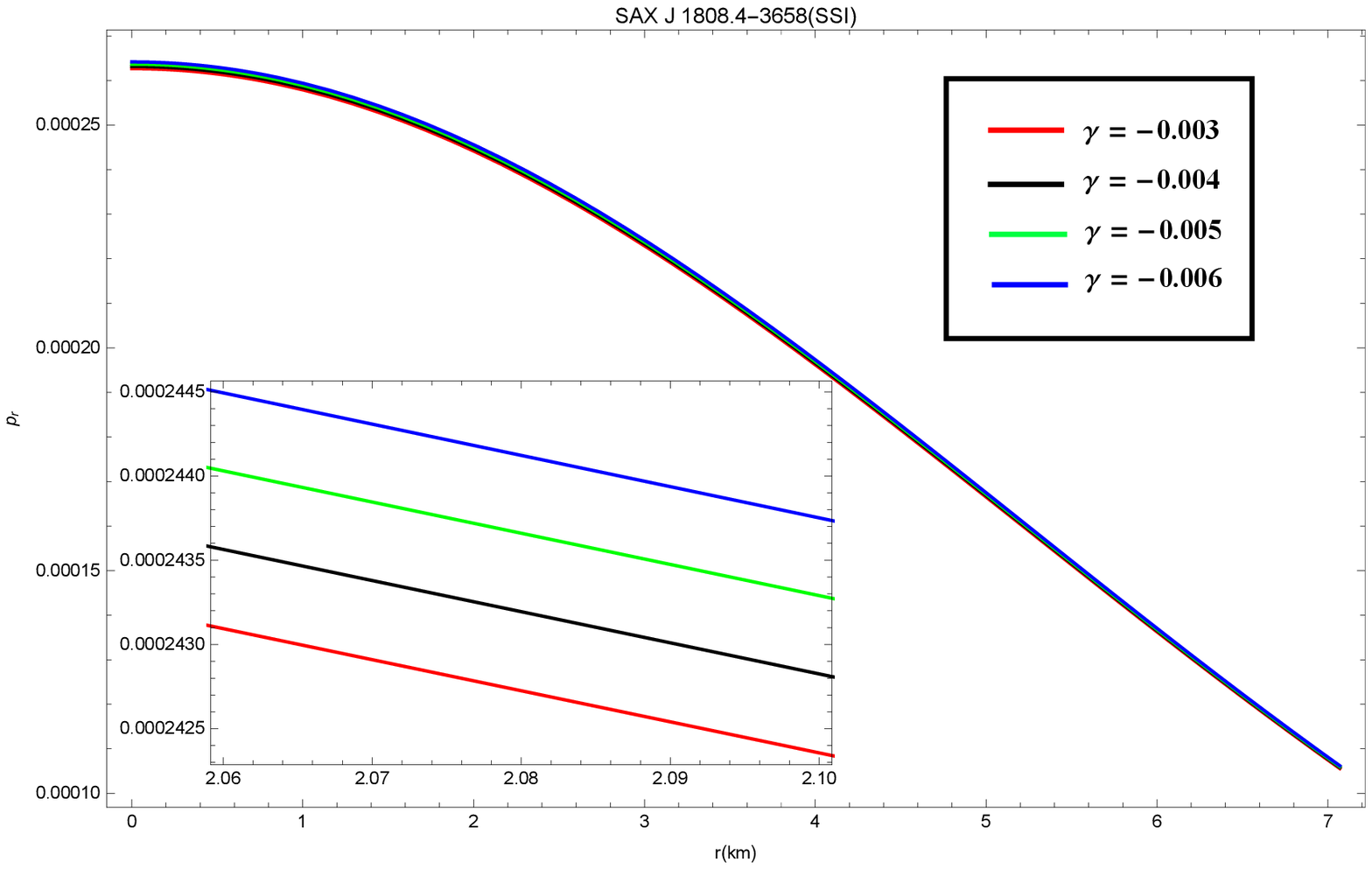}
\includegraphics[width=.50\linewidth, height=1.8in]{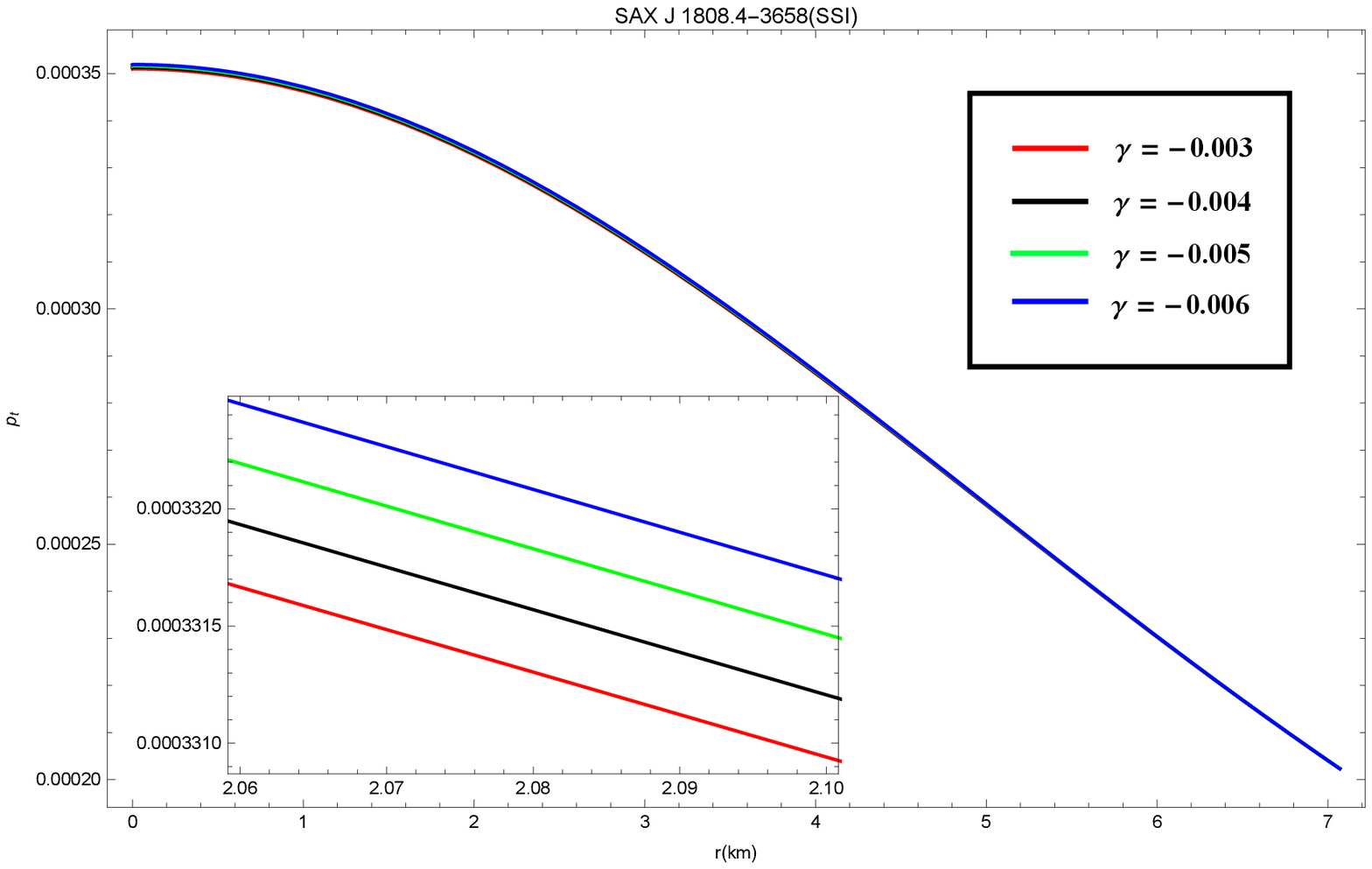}
\includegraphics[width=.49\linewidth, height=1.8in]{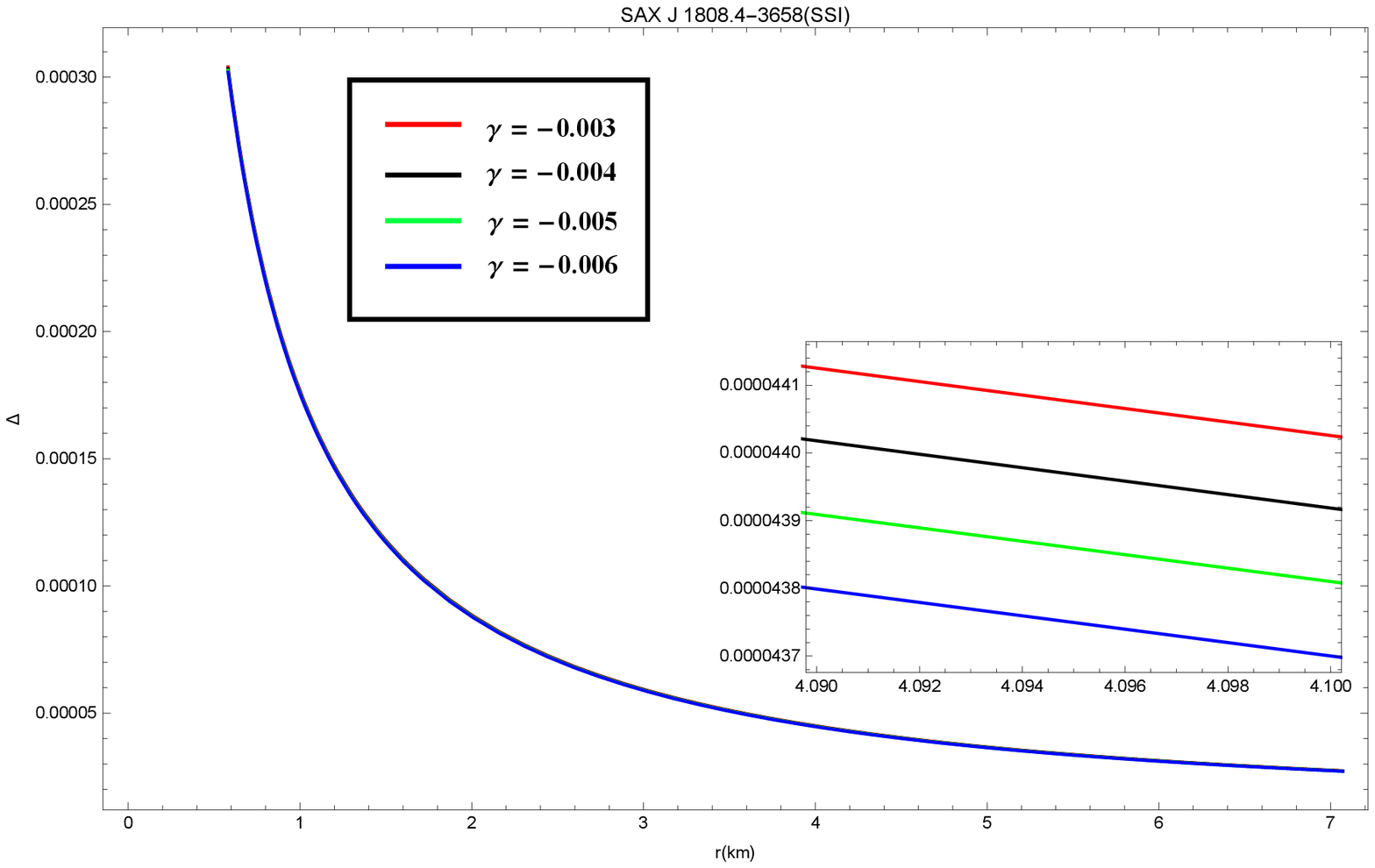}
\includegraphics[width=.50\linewidth, height=1.8in]{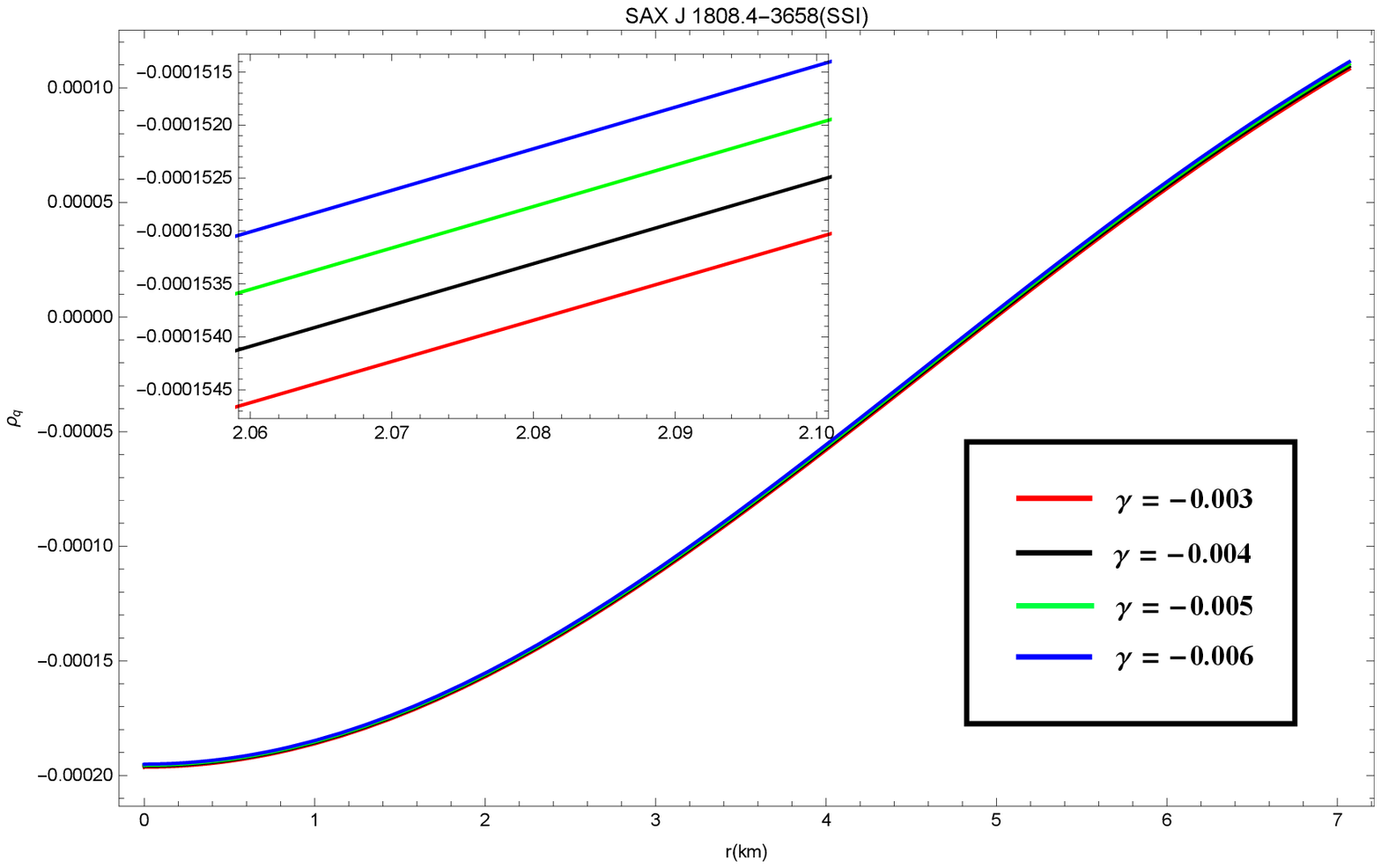}
\caption{Plots of radial pressure (upper left panel ), transverse pressure (upper right panel), anisotropic pressure (lower left panel ) and quintessence field (lower right panel) versus radius of compact star $SAX J 1808.4-3658 (SSI)$. }
\end{figure}

\begin{figure}
\includegraphics[width=.49\linewidth, height=1.8in]{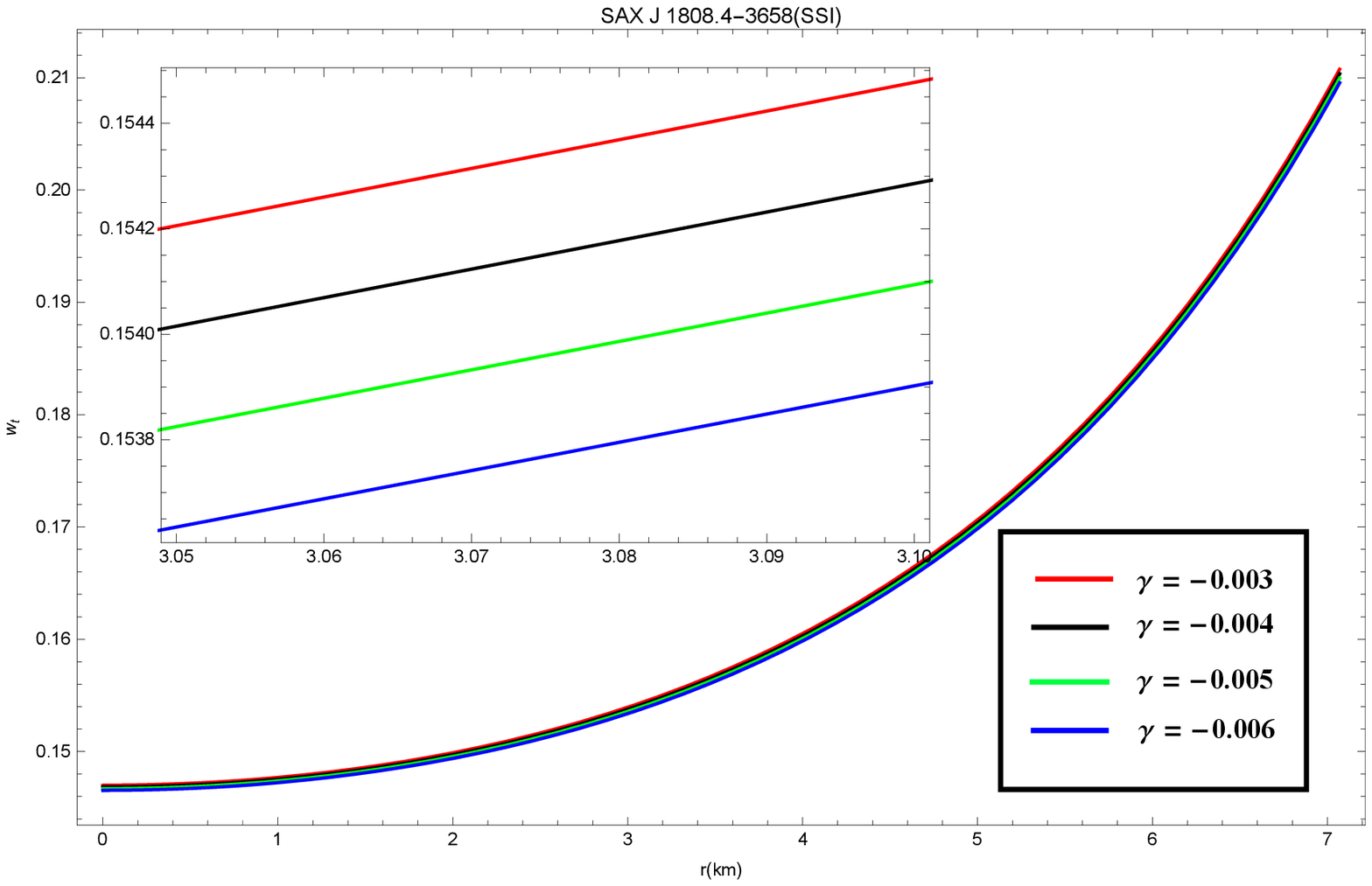}
\includegraphics[width=.49\linewidth, height=1.8in]{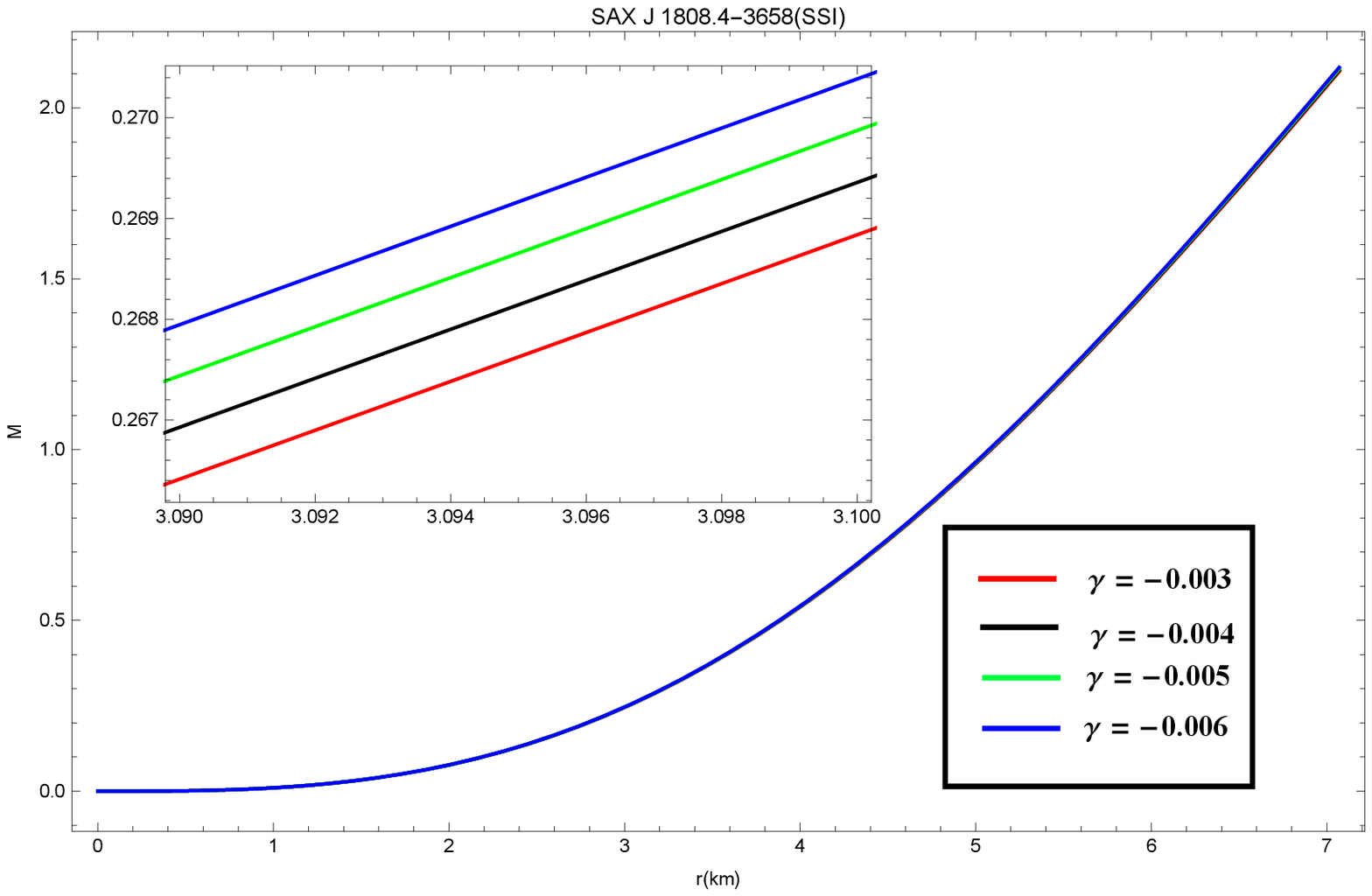}
\includegraphics[width=.49\linewidth, height=1.8in]{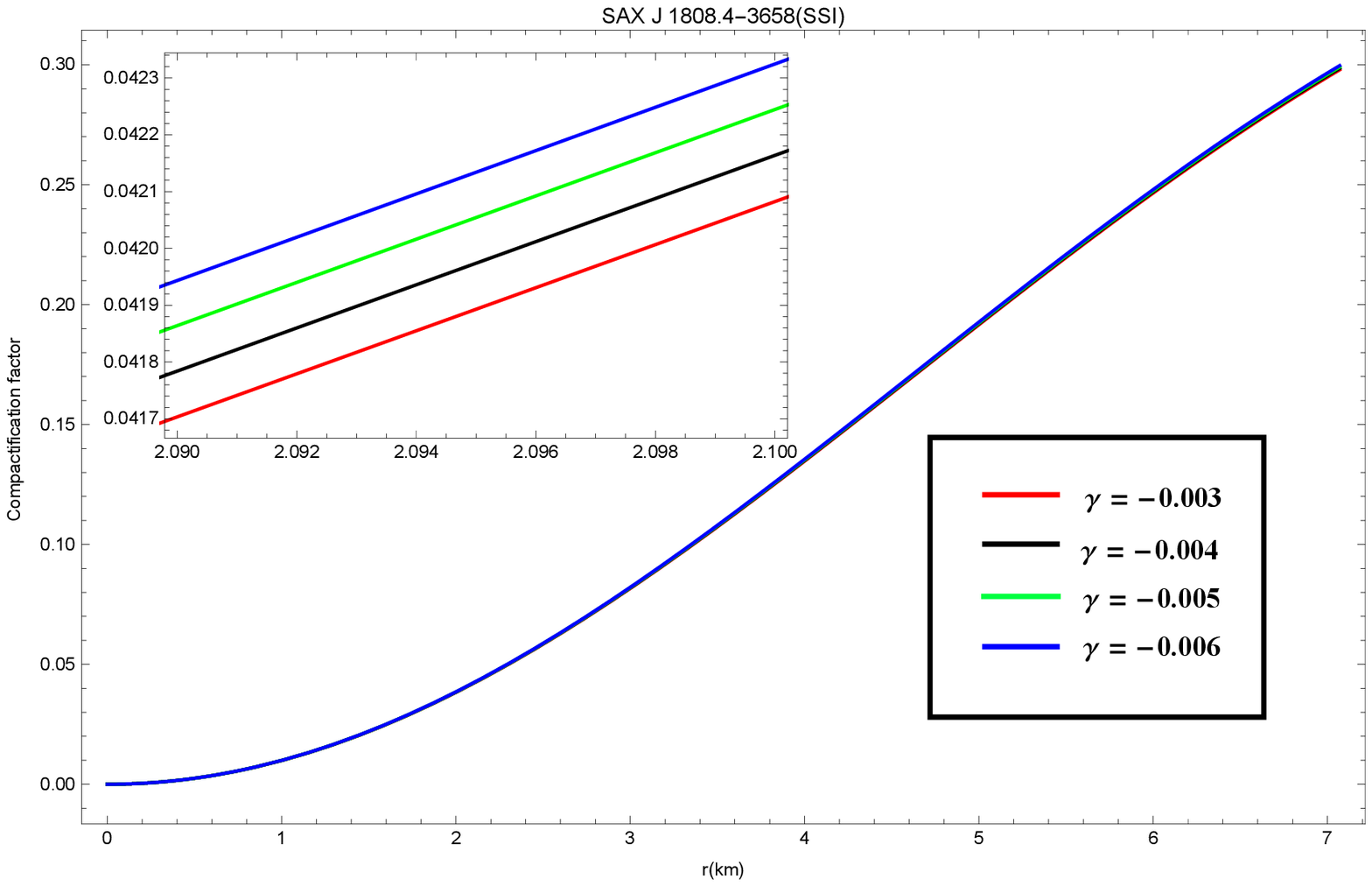}
\includegraphics[width=.49\linewidth, height=1.8in]{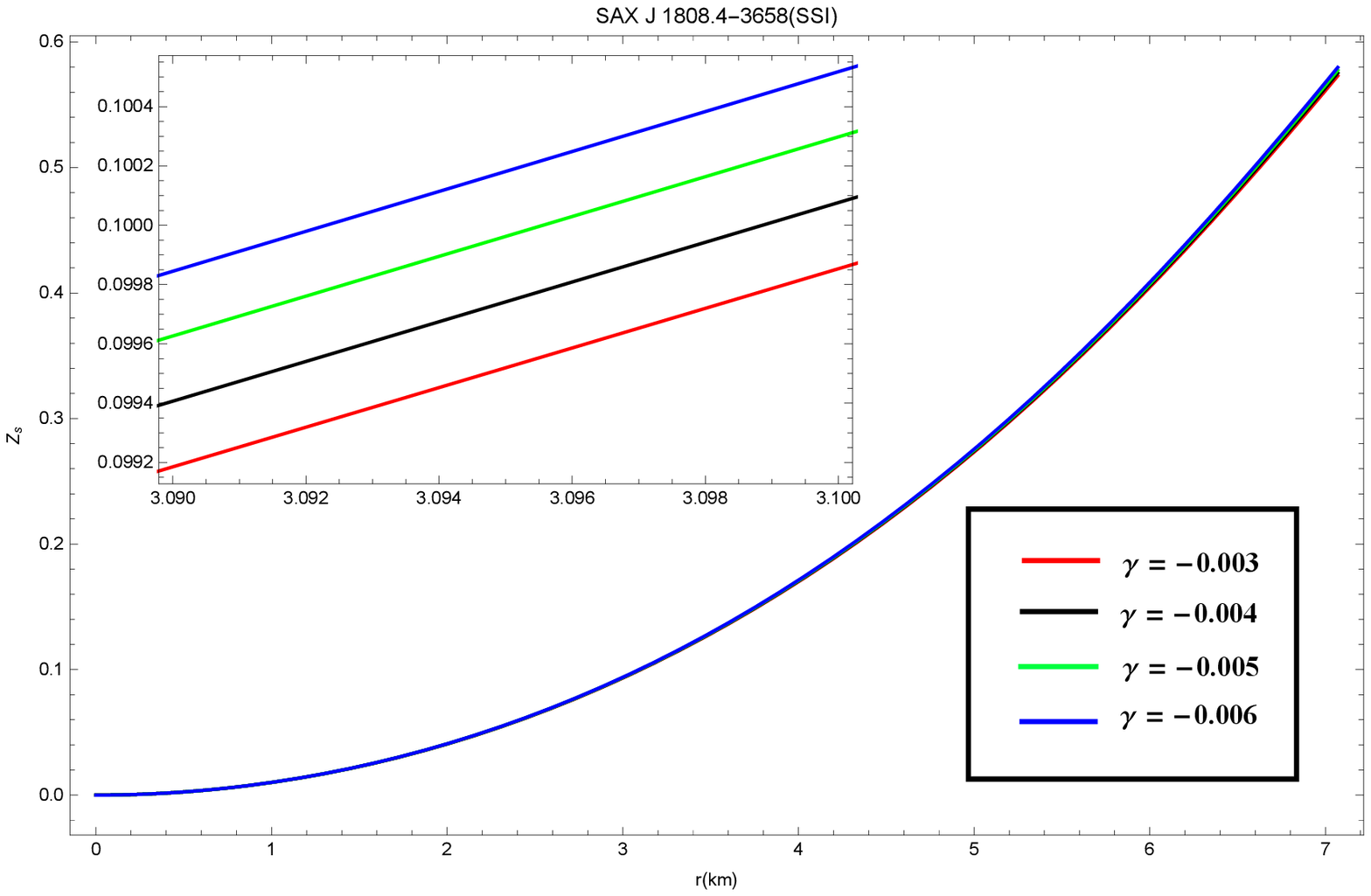}
\caption{Evolution of equation of state parameter $w_{t}$ (upper left panel ), mass function (upper right panel), compactification factor (lower left panel ) and surface redshift (lower right panel).}
\end{figure}

%..................................double Table.111...Landscape.........................................

\begin{landscape}
\begin{table}[ht]
\caption{Comparison of the parameters calculated by $GR$ and our model For Strange Star $SAX J 1808.4-3658(SSI)$}
\begin{center}
\begin{tabular}{|c|c|c|c|c|c|c|}
\hline {$\gamma$}     &{Mass}   &{Mass from}   &{Mass from}    &\textbf{$\rho_{0}$}   &\textbf{ $\rho_{R}$}   &\textbf{$p_{0}$}\\
& Standard Data& $GR$ model& our model&$(gm/cc)$&$(gm/cc)$&$(dyne/cm^{2})$\\
&($km$)&$(km)$&$(km)$&&&
\\\hline $-0.003$   & 2.116625   & 2.093970981   & 2.106385828   & $3.222050817\times10^{15}$     & $1.295264429\times10^{15}$    & $3.189830309$$\times10^{35}$
\\\hline $-0.004$   & =        & =           & 2.110458941   & $3.228281288\times10^{15}$     & $1.297769078\times10^{15}$    & $3.195998476$$\times10^{35}$
\\\hline $-0.005$   & =         &=            & 2.114500108   & $3.234462893\times10^{15}$     & $1.300254083\times10^{15}$    & $3.202118264$$\times10^{35}$
\\\hline $-0.006$   & =         &=            & 2.118509703   & $3.240596204\times10^{15}$     & $1.302719674\times10^{15}$    & $3.208190242$$\times10^{35}$
\\\hline
\end{tabular}
\end{center}

\caption{Comparison of the parameters calculated by $GR$ and our model For Strange Star $4U 1820-30$}
\begin{center}
\begin{tabular}{|c|c|c|c|c|c|c|}
\hline {$\gamma$}     &{Mass}   &{Mass from}   &{Mass from}    &\textbf{$\rho_{0}$}   &\textbf{ $\rho_{R}$}   &\textbf{$p_{0}$}\\
& Standard Data& $GR$ model& our model&$(gm/cc)$&$(gm/cc)$&$(dyne/cm^{2})$\\
&($km$)&$(km)$&$(km)$&&&
\\\hline $0.007$    & 3.31875   & 3.383132691  & 3.334404446   & $1.981906288\times10^{15}$     & $6.659205128\times10^{14}$    & $1.962087225$$\times10^{35}$
\\\hline $0.008$    & =         & =            & 3.327213146   & $1.977631917\times10^{15}$     & $6.644843240\times10^{14}$    & $1.957855598$$\times10^{35}$
\\\hline $0.009$    & =         & =            & 3.319962167   & $1.973322073\times10^{15}$     & $6.630362167\times10^{14}$    & $1.953588853$$\times10^{35}$
\\\hline $0.01$     & =         & =            & 3.312650763   & $1.968976315\times10^{15}$     & $6.615760418\times10^{14}$    & $1.949286552$$\times10^{35}$
\\\hline
\end{tabular}
\end{center}
\caption{Comparison of the parameters calculated by $GR$ and our model For Strange Star $Vela X-12$ }
\begin{center}
\begin{tabular}{|c|c|c|c|c|c|c|}
\hline {$\gamma$}     &{Mass}   &{Mass from}   &{Mass from}    &\textbf{$\rho_{0}$}   &\textbf{ $\rho_{R}$}   &\textbf{$p_{0}$}\\
& Standard Data& $GR$ model& our model&$(gm/cc)$&$(gm/cc)$&$(dyne/cm^{2})$\\
&($km$)&$(km)$&$(km)$&&&
\\\hline $-0.016$    & 2.61075   & 2.093970981  & 2.521505015   & $1.285043654\times10^{15}$     & $6.133864535\times10^{14}$   & $1.272193218$$\times10^{35}$
\\\hline $-0.017$    & =         & =            & 2.605638649   & $1.287239319\times10^{15}$     & $6.144345044\times10^{14}$    & $1.274366926$$\times10^{35}$
\\\hline $-0.018$    & =         & =            & 2.610049962   & $1.289418599\times10^{15}$     & $6.154747340\times10^{14}$    & $1.276524413$$\times10^{35}$
\\\hline $-0.019$    & =         & =            & 2.614428476   & $1.291581675\times10^{15}$     & $6.165072296\times10^{14}$    & $1.278665858$$\times10^{35}$
\\\hline
\end{tabular}
\end{center}

\end{table}
\end{landscape}
%..................................double Table.333...Landscape.........................................

\begin{landscape}
\begin{table}[ht]

\caption{Comparison of the parameters calculated by $GR$ and our model For Strange Star $PSR J1614-2230$}
\begin{center}
\begin{tabular}{|c|c|c|c|c|c|c|}
\hline {$\gamma$}     &{Mass}   &{Mass from}   &{Mass from}    &\textbf{$\rho_{0}$}   &\textbf{ $\rho_{R}$}   &\textbf{$p_{0}$}\\
& Standard Data& $GR$ model& our model&$(gm/cc)$&$(gm/cc)$&$(dyne/cm^{2})$\\
&($km$)&$(km)$&$(km)$&&&
\\\hline $-0.010$    & 2.90575   & 2.845190244  & 2.899905441   & $1.373589358\times10^{15}$     & $5.985782365\times10^{14}$    & $1.359853464$$\times10^{35}$
\\\hline $-0.011$    & =         & =            & 2.905146360   & $1.376071808\times10^{15}$     & $5.996600304\times10^{14}$    & $1.362311090$$\times10^{35}$
\\\hline $-0.012$    & =         & =            & 2.910347272   & $1.378535308\times10^{15}$     & $6.007335664\times10^{14}$    & $1.364749955$$\times10^{35}$
\\\hline $-0.013$    & =         & =            & 2.915508634   & $1.380980075\times10^{15}$     & $6.017989386\times10^{14}$    & $1.367170274$$\times10^{35}$
\\\hline
\end{tabular}
\end{center}
\caption{Comparison of the parameters calculated by $GR$ and our model For Strange Star $SAX J 1808.4-3658(SSI)$}
\begin{center}
\begin{tabular}{|c|c|c|c|c|c|c|}
\hline {$\gamma$}     &{$\frac{M}{R}$}   &{$\frac{M}{R}$ from}   &{$\frac{M}{R}$ from}    &{$\frac{2M}{R}$ from}   &{( $\rho_{R}$/$\rho_{0}$) from}  &{$Z_{s}$ (Max Value)}\\
& Standard Data& $GR$ model& our model&our model&our model & from our model
\\\hline $-0.003$ &0.2993811881 &0.2961769421 & 0.2979329318   & 0.595865863$<\frac{8}{9}$      & 0.402    & 0.573030821
\\\hline $-0.004$ &=            & =           & 0.2985090440   & 0.597018088$<\frac{8}{9}$      & 0.402    & 0.575278057
\\\hline $-0.005$ &=            & =           & 0.2990806376   & 0.598161275$<\frac{8}{9}$      & 0.402    & 0.577517212
\\\hline $-0.006$ &=            & =           & 0.2996477656   & 0.599295531$<\frac{8}{9}$      & 0.402    & 0.579748338
\\\hline
\end{tabular}
\end{center}

\caption{Comparison of the parameters calculated by $GR$ and our model For Strange Star $4U 1820-30$}
\begin{center}
\begin{tabular}{|c|c|c|c|c|c|c|}
\hline {$\gamma$}     &{$\frac{M}{R}$}   &{$\frac{M}{R}$ from}   &{$\frac{M}{R}$ from}    &{$\frac{2M}{R}$ from}   &{( $\rho_{R}$/$\rho_{0}$) from}  &{$Z_{s}$ (Max Value)}\\
& Standard Data& $GR$ model& our model&our model&our model & from our model
\\\hline $0.007$  & 0.3318750000 & 0.3383132691 & 0.3334404446  &0.6668808892$<\frac{8}{9}$      & 0.3360000000    & 0.732607642
\\\hline $0.008$  & =            & =            & 0.3327213146  &0.6654426292$<\frac{8}{9}$      & 0.3359999999    & 0.728879396
\\\hline $0.009$  & =            & =            & 0.3319962167  &0.6639924334$<\frac{8}{9}$      & 0.3360000001    & 0.725144474
\\\hline $0.01$   & =            & =            & 0.3312650763  &0.6625301526$<\frac{8}{9}$      & 0.3359999999    & 0.721402830
\\\hline
\end{tabular}
\end{center}

\end{table}
\end{landscape}
%.................................................end double Table.3333............................................

%..................................double Table.4444444444....Landscape.........................................
\begin{landscape}
\begin{table}[ht]
\caption{Comparison of the parameters calculated by $GR$ and our model For Strange Star $Vela X-12$ }
\begin{center}
\begin{tabular}{|c|c|c|c|c|c|c|}
\hline {$\gamma$}     &{$\frac{M}{R}$}   &{$\frac{M}{R}$ from}   &{$\frac{M}{R}$ from}    &{$\frac{2M}{R}$ from}   &{( $\rho_{R}$/$\rho_{0}$) from}  &{$Z_{s}$ (Max Value)}\\
& Standard Data& $GR$ model& our model&our model&our model & from our model
\\\hline $-0.016 $ & 0.2613363363& 0.2524029044 & 0.2524029044   &0.5048058088$<\frac{8}{9}$      &0.4773273278   & 0.444519091
\\\hline $-0.017 $ & =           & =            & 0.2608246896   &0.5216493792$<\frac{8}{9}$      &0.4773273278   & 0.445861951
\\\hline $-0.018 $ & =           & =            & 0.2612662625   &0.5225325250$<\frac{8}{9}$      &0.4773273276   & 0.447198499
\\\hline $-0.019 $ & =           & =            & 0.2617045521   &0.5234091042$<\frac{8}{9}$      &0.4773273278   & 0.448528782
\\\hline
\end{tabular}
\end{center}

\caption{Comparison of the parameters calculated by $GR$ and our model For Strange Star $PSR J1614-2230$}
\begin{center}
\begin{tabular}{|c|c|c|c|c|c|c|}
\hline {$\gamma$}     &{$\frac{M}{R}$}   &{$\frac{M}{R}$ from}   &{$\frac{M}{R}$ from}    &{$\frac{2M}{R}$ from}   &{( $\rho_{R}$/$\rho_{0}$) from}  &{$Z_{s}$ (Max Value)}\\
& Standard Data& $GR$ model& our model&our model&our model & from our model
\\\hline $-0.010$    & 0.2821116505   & 0.2762320625   & 0.2815442176   & 0.5630884352$<\frac{8}{9}$     & 0.4357766992    & 0.512875639
\\\hline $-0.011$    & =              & =              & 0.2820530447   & 0.5641060894$<\frac{8}{9}$     & 0.4357766992    & 0.514640618
\\\hline $-0.012$    & =              & =              & 0.2825579876   & 0.5651159752$<\frac{8}{9}$     & 0.4357766993    & 0.516398244
\\\hline $-0.013$    & =              & =              & 0.2830590907   & 0.5661181814$<\frac{8}{9}$     & 0.4357766991    & 0.518148567
\\\hline
\end{tabular}
\end{center}

\end{table}
\end{landscape}

%.................................................end double Table.44444.........................................
\section{Conclusions}

In the present work, motivated by the observational evidences of the rapid expansion of our cosmos, we have used the quintessence dark energy having characteristic parameter $\omega_{q}$ satisfying the constraint $-1<\omega_{q}<-\frac{1}{3}$, to develop a compact star model in the Rastall framework. To demonstrate the physical acceptance of our propounded model we have performed various physical tests with the help of different parameters. For the sake of clarity in our findings we also present the graphical analysis (\textbf{Figs. 1-4}) of the physical parameters involved in our model. A comparative study is also performed, in which we compare our obtained results with that of $GR$ model and observational data as well, and observed that the presented model is more compatible with the observational data than $GR$ model. Now we summarize our findings in the following points:
\begin{itemize}
 \item
 \textbf{Density and pressure:} Among the necessary conditions for a physical model to be viable,  pressure (radial and transverse) and energy density should remains positive throughout the stellar interior and gain their corresponding maximum values at the center, then decrease monotonically as $r$ increases towards the surface. For the present model, \textbf{Figs. 1} and \textbf{3} endorsed the good behavior of the energy density and pressure. From our present model we have computed the numerical values of density at the center and surface and the value of pressure at the center, using equations (\ref{s1})-(\ref{s3}) respectively. One can note from the \textbf{Tables 2-5} that central density of different compact stars ($\sim 10^{15} gm/cc$) are beyond the nuclear density ($\sim 10^{14} gm/cc$).

\item  \textbf{Energy conditions:} We present the graphical analysis of all the energy conditions given in inequalities (\ref{r13a})-(\ref{r13d}). Evidently it can be seen from  \textbf{Figs. 1} that all the energy conditions mentioned above are satisfied for our system.

\item  \textbf{Stability:} To illustrate the present model is stable we have analyzed both the conditions, i.e., Herrera cracking condition (also called causality condition) \cite{42} and the condition for potentially stable region \cite{42,43}, with the aid of plot. \textbf{Fig. 2} indicate that both the conditions, i.e., $\nu^{2}_{sr}\leq1$,  $\nu^{2}_{st}\leq1$ and $|\nu^{2}_{st}-\nu^{2}_{sr}|\leq1$ are satisfied and hence our presented model is physically plausible and potentially stable with in the entire fluid sphere with anisotropic configuration.

\item  \textbf{Anisotropic Pressure:} We have observed, in connection with graphical representations (\textbf{Fig. 3}), that the measure of anisotropic pressure $\Delta=\frac{2}{r}(p_{t}-p_{r})$ is positive throughout the region of the compact stars considered in this study. As discussed earlier that for $\Delta>0$, i.e., $p_{t}>p_{r}$, the anisotropic force is outward directed and for $\Delta<0$ this force is directed inward. In the present case $\Delta>0$ indicating that the anisotropic stress is repulsive and hence support the construction of more compact object.

\item  \textbf{Equation of state:} Corresponding to the radial and tangential directions, we have examined the equation of state ($EoS$) parameters $w_{r}$ and $w_{t}$. Using the relations for $w_{r}$ and $w_{t}$ with the aid of \textbf{Fig. 4} (upper left panel), we have observed that the value of these parameters remain positive inside the stellar objects and also less than 1. Which is an other evidence for good behavior of the model.

\item  \textbf{Mass function:} We have plotted the mass function in \textbf{Fig. 4} (upper right panel) and note that as $r\rightarrow0$, $M\rightarrow0$. Also it is evident from figure that mass function is monotonically increasing toward the surface. We have computed the numerical values of the masses, using our proposed model for different compact stars and shown in \textbf{Tables 2-5}. It is eminently clear that our computed values of masses (for the chosen values of the Rastall dimensionless parameter) for different stellar objects are more closer to the observational data than the $GR$ model proposed in \cite{1}.

\item  \textbf{ Mass-radius ratio ( Compactification factor):}  In \cite{55a} Buchdahl introduced the upper limit for the mass to radius ratio which is $\frac{M}{R}<\frac{4}{9}$ or equivalently $\frac{2M}{R}<\frac{8}{9}$. Also a generalized version of the Buchdahl limit is given in \cite{55b}. In this sequel we have defined the compactification factor $u(R)$ as the ratio of the mass $M(r)$ and radius $R$. The profile of $u(R)$ has been shown in \textbf{Fig. 4} (lower left panel) depicting the increasing nature of this factor from center to surface. Also we have computed the value of compactification factor using our model given in \textbf{Tables 6-9}. We have observed from \textbf{Tables 6-9} that the Buchdahl condition is fulfilled by our model.

\item \textbf{Surface redshift:} The profile of the surface redshift has been shown in \textbf{Fig. 4} (lower right panel). It is observed that $Z_{s}$ is zero at the center and increases with the increase in $r$. The maximum value of the surface redshift has been computed from our model for different compact stars and given in \textbf{Tables 6-9}. In view of the discussion in \cite{55d}-\cite{55g} about the upper limit of the surface redshift, we argued that the maximum numerical value for surface redshift computed from our present model $Z_{s}=0.733$ is in good concurrence.
\end{itemize}
\section*{Acknowledgment}

We{\color{white}i}appreciate{\color{white}i}the financial support from{%
\color{white}i}HEC, Islamabad, Pakistan{\color{white}i}under NRPU project{%
\color{white}i}with grant{\color{white}i}number 20-4059/NRPU/R \&
D/HEC/14/1217. 

\vspace{0.25cm}

\begin{thebibliography}{40}
\bibitem{1} P. Bhar, Astrophys. Space Sci. \textbf{356}, 309 (2015).
%\bibitem{2} R. Ruderman, Rev. Astron. Astrophys. \textbf{10}, 427 (1972).




\bibitem{22} P. Rastall, Phys. Rev. D \textbf{6}, 3357 (1972).

\bibitem{55} I. Licata, H. Moradpour and C. Corda, Int. J. Geom. Methods Mod. Phys. \textbf{14}, 1730003 (2017).
\bibitem{57} M. Capone, V. F. Cardone and M. L. Ruggiero, j. Phys. Conf. Ser. \textbf{222}, 012012 (2010).
\bibitem{58} R. Kumar, B. P. Singh, M. S. Ali and S. G. Ghosh, arXiv:1712.09793 [gr-qc].
\bibitem{60} F. Darabi, K. Atazadeh and Y. Heydarzade, Eur. Phys. J. Plus \textbf{133}, 259 (2018).
\bibitem{61} J. C. Fabris, O. F. Piattella, D. C. Rodrigues, C. E. M. Batista and M. H. Daouda, Int. J. Mod. Phys. Con. Ser. \textbf{18}, 67 (2012).
\bibitem{44} Y. Heydarzade, H. Hadi, C. Corda and F. Darabi, Phys. Lett. \textbf{B776}, 457(2018).



\bibitem{23} P. Rastall, Can. J. Phys. \textbf{54}, 66 (1976).

\bibitem{23b} Y. Heydarzade and F. Darabi, Phys. Lett. B \textbf{771}, 365-373 (2017).
\bibitem{23c} Y. Heydarzade, H. Moradpour and F. Darabi, Can. J. Phys., \textbf{95}, 1253-1256 (2017).
\bibitem{23d} R. Kumar and S. G. Ghosh, Eur.Phys. J. C \textbf{78}, 750 (2018).
\bibitem{23e} M. S. Ma and R. Zhao, Eur. Phys. J. C \textbf{77}, 629 (2017).
\bibitem{40} I. P. Lobo, H. Moradpour, J. P. Morais Graca and I. G. Salako, Int. J. Mod. Phys D \textbf{27}, 1850069 (2018).
\bibitem{23a} J. C. Fabris, O. F. Piattella, D. C. Rodrigues, C. E. M. Batista and M. H. Daouda, Int. J. Mod. Phys. Conf. Ser. \textbf{18}, (2012)67-76.
\bibitem{48} H. Moradpour and I.G. Salako, Adv. High Energy Phys. \textbf{2016}, 3492796(2016).
\bibitem{54} H. Moradpour, N. Sadeghnezhad and S.H. Hendi, Can. J. Phys. \textbf{95}, 1257 (2017).
\bibitem{45} H. Moradpour, A. Bonilla, E. M. C. Abreu and J. A. Neto, Phys. Rev. D \textbf{96}, 123504 (2017).
\bibitem{46} H. Moradpour, Y. Heydarzade, F. Darabi and I. G. Salako, Eur. Phys. J. C \textbf{77}, 259 (2017).
\bibitem{47} F. Darabi, H. Moradpour, I. Licata, Y. Heydarzade and C. Corda, Eur. Phys. J. C \textbf{78}, 25 (2018).
\bibitem{470} M. Visser, Phys. Lett. B \textbf{782}, 83 (2018).
\bibitem{47a} S. Hansraj, A. Banerjee and P. Channuie, arXiv: 1805.00003v1 [gr-qc].
\bibitem{47b} S. Hansraj and A. Banerjee, arXiv: 1807.00812v1 [gr-qc].


\bibitem{3}R. L. Bowers and E. P. T. Liang, Class. Astrophys. J. \textbf{188}, 657 (1974).

\bibitem{4} L. Herrera and N. O. Santos, Phys. Report. \textbf{286}, 53 (1997).
\bibitem{a4} S. K. Maurya, A. Banerjee, and S. Hansraj, Phys. Rev. D \textbf{97}, 044022 (2018).
\bibitem{a5} S.K. Maurya, S. Ray, S. Ghosh, S. Manna and T. T. Smitha, Annal. Phys. \textbf{395}, 152 (2018).
\bibitem{4a} L. Herrera, Phys. Lett. A \textbf{165}, 206 (1992).

\bibitem{4b} L. Herrera, A. Di Prisco and J. Ibanez, Phys. Rev. D \textbf{84}, 107501 (2011).

\bibitem{4c} L. Herrera, J. Ospino and A. Di Prisco, Phys. Rev. D \textbf{77}, 027502 (2008).

\bibitem{4d} L. Herrera, N. O. Santos and A. Wang, Phys. Rev. D \textbf{78}, 084026 (2008).

\bibitem{5a} M. Sharif and G. Abbas, J. Phys. Soc. Jpn. \textbf{82}, 034006 (2013).

\bibitem{5b} M. Sharif and G. Abbas, Chin. Phys. B \textbf{22}, 030401 (2013).

\bibitem{5c} M. Sharif and G. Abbas, Eur. Phys. J. Plus \textbf{28}, 10 (2013).

\bibitem{5} B. V. Ivanov, Phys. Rev. D \textbf{65}, 104011 (2002).

\bibitem{12} F. Rahaman, R. Maulick , A. K. Yadav, S. Ray and R. Sharma, Gen. Relativ.Gravit. \textbf{44}, 107 (2012).

\bibitem{13} M. Kalam, F. Rahaman, S. Ray, Sk. M. Hossein, I. Karar and J. Naskar, Eur.Phys. J. C \textbf{72}, 2248 (2012).

\bibitem{16} S.K. Maurya, Y. K. Gupta, S. Ray and D. Deb, Eur. Phys. J. C \textbf{76}, 693(2016).

\bibitem{17} S. K. Maurya, D. Deb, S. Ray and P. K. F. Kuhfittig, eprint arXiv:1703.08436.


\bibitem{41} K. D. Krori and J. Barua, J. Phys. A.: Math. Gen. \textbf{8}, 508 (1975).
%\bibitem{41a} M. Zubair and G. Abbas, Astrophys. Space Sci. \textbf{361}, 27 (2016).
%\bibitem{41b} M. Z. Bhatti, M. Sharif, Z. Yousaf and M. Ilyas, Int. J. Mod. Phys. D \textbf{27}, 1850044 (2018)

\bibitem{42} L. Herrera, Phys. Lett. A \textbf{165}, 206 (1992).
\bibitem{43} H. Andreasson, Commun. Math. Phys. \textbf{288}, 715 (2009).


\bibitem{51} A. F. Santos and S.C. Ulhoa, Mod. Phys. Lett. A \textbf{30}, 1550039 (2015).
\bibitem{49} F. F. Yuan and P. Huang, Class. Quantum Gravity \textbf{34}, 077001 (2017).

\bibitem{55a} H.A. Buchdahl, Phys. Rev. \textbf{116}, 1027 (1959).
\bibitem{55b} M.K. Mak, T. Harko, Proc. R. Soc. A \textbf{459}, 393 (2003).

\bibitem{55d} N. Straumann, General Relativity and Relativistic Astrophysics (Springer Verlag, Berlin, (1984).
\bibitem{55e} C.G. Bohmer, T. Harko, Gen. Relativ. Gravit. \textbf{39}, 757 (2007).
\bibitem{55f} C.G. Bohmer, T. Harko, Class. Quantum Gravit. \textbf{23}, 6479 (2006).
\bibitem{55g} B.V. Ivanov, Phys. Rev. D \textbf{65}, 104001 (2002).




\end{thebibliography}
\end{document}